\documentclass[epj]{svjour}
\usepackage{graphicx}
\usepackage{units,amsmath,amssymb}
\usepackage[pdfborder={0 0 0}]{hyperref}
\usepackage{rotating}

\begin{document}

\newcommand{\figref}[1]{Fig.~\ref{#1}}
\newcommand{\tabref}[1]{Tab.~\ref{#1}}
\newcommand{\secref}[1]{Section \ref{#1}}
\newcommand{\appref}[1]{Appendix \ref{#1}}
\title{Introducing an interface between FeynRules and WHIZARD}

\author{%
  Neil D. Christensen\thanks{\email{neilc@pitt.edu}}\inst{,1}\and
  Claude Duhr\thanks{\email{claude.duhr@durham.ac.uk}}\inst{,2}\and
  Benjamin Fuks\thanks{\email{benjamin.fuks@iphc.cnrs.fr}}\inst{,3}\and
  J\"urgen Reuter\thanks{\email{juergen.reuter@desy.de}}\inst{,4,5,6}\and
  Christian Speckner\thanks{\email{christian.speckner@physik.uni-freiburg.de}}\inst{,5}
}
\institute{%
PITTsburgh Particle physics, Astrophysics and Cosmology Center,
Department of Physics and Astronomy, University of Pittsburgh,
Pittsburgh, PA 15260, USA \and
Institute for Particle Physics Phenomenology, University of Durham,
Durham, DH1 3LE, U.K. \and
Institut Pluridisciplinaire Hubert Curien/D\'epartement Recherches Subatomiques,
Universit\'e de Strasbourg/CNRS-IN2P3, 
23 Rue du Loess, F-67037 Strasbourg, France \and
DESY Theory Group, Notkestra{\ss}e 85, D-22603 Hamburg, Germany \and
Albert-Ludwigs-Universit\"at Freiburg, Physikalisches Institut,
Hermann-Herder-Stra\ss{}e 3, 79104 Freiburg, Germany \and
University of Edinburgh\mbox{,} School of Physics and Astronomy,
JCMB, The King's Buildings, Mayfield Road, Edinburgh EH9 3JZ, U.K.
}

\date{February 2012}

\headnote{\begin{footnotesize}
PITT PACC 12002, \quad DESY 12-020, \quad
FR-PHENO-2010-030, \quad IPHC-PHENO-10-03, 
\quad IPPP/10/80, DCPT/10/160, \quad MADPH-10-1562,
\quad EDINBURGH-2010-25
\end{footnotesize}}

\abstract{
While Monte Carlo event generators like {\sc Whizard} have become indispensable tools
in studying the impact of new physics on collider observables over the last
decades, the implementation of new models in such packages has remained a rather
awkward and error-prone process. Recently, the {\sc FeynRules} package was introduced
which greatly simplifies this process by providing a single unified model format
from which model implementations for many different Monte Carlo codes can be
derived automatically. In this note, we present an interface which extends
{\sc FeynRules} to provide this functionality also for the {\sc Whizard} package, thus making
{\sc Whizard}'s strengths and performance easily available to model builders.
}


\maketitle

\section{Introduction}

The Standard Model (SM) of particle physics provides a successful description of all
experimental high-energy data to date. 
However, despite its success, many fundamental questions remain unanswered, 
such as the origin of electroweak symmetry breaking, the nature of
neutrino masses, the large hierarchy between the electroweak and the
Planck scales and the origins of dark matter and the cosmological constant. 
Attempts to address these questions have led to a wide range of new physics
theories, most of them predicting new phenomena at the TeV scale. 
The Large Hadron Collider (LHC) at CERN is currently probing this
scale and will hopefully make it possible to discover, constrain
and/or exclude some of the theories beyond the Standard Model (BSM).

Monte Carlo simulations of the particle collisions to be observed at the LHC will
play a key role in the exploration of the weak scale, both from the
theoretical and experimental sides. While proper modelling of the strong
interactions, including parton showering, fragmentation and hadronization, is
essential in order to achieve a realistic description of the event distributions
at the LHC, BSM signals are expected to show up predominantly in the underlying
hard interaction. Therefore, a lot of effort has been put into the development of 
multi-purpose matrix-element and parton-level event generators such as
{\sc CompHep}/{\sc CalcHep}\ \cite{Pukhov:1999gg,Boos:2004kh,Pukhov:2004ca},
{\sc MadGraph}/{\sc MadEvent}\
\cite{Stelzer:1994ta,Maltoni:2002qb,Alwall:2007st,Alwall:2008pm,Alwall:2011uj}, 
{\sc Sherpa} \cite{Gleisberg:2008ta} and {\sc Whizard} \cite{Moretti:2001zz,Kilian:2007gr},
which allow, at least in principle, the generation of parton-level events for a large 
class of Lagrangian-based quantum field theory models.

In practice, however, the implementation of a BSM model into any of these tools can be a 
tedious and error-prone task. Not only might a model contain hundreds, if not thousands,
of interaction vertices which need to be encoded in a format suitable for the
generator in question, but in addition, each code follows its own
format conventions
which need to be respected.
To improve this situation, a framework based on the {\sc FeynRules} package
\cite{Christensen:2008py}, addressing not only the implementation but also
the validation of new physics models in multi-purpose
matrix-element and event generators has recently been
proposed~\cite{Christensen:2009jx}  and its virtue illustrated in the
context of the programs {\sc CompHep}/{\sc CalcHep}, {\sc MadGraph}/{\sc Mad\-Event}
and {\sc Sherpa}.  

In this paper we present a new interface from {\sc FeynRules} to the event
generator {\sc Whizard}. It follows the same approach as the already 
existing interfaces, and allows to export any model implemented in {\sc FeynRules} 
directly to {\sc Whizard}, extending thus the aforementioned framework to the
{\sc Whizard} package. The paper is organized as follows: after giving a short
overview of the {\sc FeynRules} and {\sc Whizard} packages in Section
\ref{sec:packages}, we discuss the usage, features and limitations of the new
interface in Section \ref{sec:WO_marries_FR}. Finally, we give a few
examples of using the interface in Section \ref{sec:usageex} before concluding in
Section \ref{sec:conclusions}. The Appendix contains an exhaustive list of the
interface options (Appendix \ref{app:interface-options}) and a selection of numerical
results from the interface validation (Appendix \ref{app:validation-tables}).


\section{The {\sc Whizard} and {\sc FeynRules} packages}
\label{sec:packages}

\subsection{FeynRules}

{\sc FeynRules} is a {\sc Mathematica}\footnote{
{\sc Mathematica} is a registered trademark of Wolfram Research Inc.}
package that allows to derive
Feynman rules from any perturbative quantum field theory-based Lagrangian 
in an automated way. The input provided by the user is threefold and consists 
of the Lagrangian defining the model, together with the definitions of
all the 
particles and parameters that appear in the model. Up to now, the public 
release of {\sc FeynRules} supports scalar, vector,
fermion (Dirac, Majorana and Weyl~\cite{Butterworth:2010ym}) and spin-two fields, as well as Faddeev-Popov
ghosts, while recently, superfields have also been included~\cite{Duhr:2011se}.
Once this information is provided, {\sc FeynRules} can perform basic checks 
on the sanity of the implementation (hermiticity, normalization of the quadratic
terms, \ldots), and finally computes all the interaction vertices associated 
with the model and store them in an internal format for later processing. 

After the Feynman rules have been obtained, {\sc FeynRules} can export 
the interaction vertices to various matrix-element generators by means 
of interfaces provided by the package \cite{Christensen:2009jx}. 
The current public release contains interfaces to {\sc CalcHep/CompHep}, 
{\sc FeynArts/FormCalc}, {\sc GoSam}~\cite{Cullen:2011ac,Degrande:2011ua}, 
{\sc MadGraph/MadEvent} (both 4 and 5), {\sc Sherpa} and
the newly developed interface to {\sc Whizard} which we present in
this note.

Since the matrix-element generators very often have color and/or Lorentz
structures hardcoded, the different interfaces check whether all the vertices
are compliant with the structures supported by the corresponding
matrix-element generators, and discard them in the case they are not supported.
The output of an interface consists of a set of files organized in a single
directory which can be installed into the relevant matrix-element
generator and used as any other built-in models.

\subsection{{\sc Whizard}}
\label{app:whizard}

{\sc Whizard} \cite{Kilian:2007gr} is a program package for the automatic and
efficient generation of unweighted parton-level events for multileg tree-level
processes in both the Standard Model and a large number of BSM models. 
While the original version of {\sc Whizard (1.xx)} was geared towards
ILC physics by offering elaborate options for controlling the beam
setup, the new version {\sc Whizard 2.x.x} has been redesigned as a
hadron collider tool vastly extending but also including the (ILC)
features of version 1 (it now also includes both a $k_T$-ordered and an
analytic parton shower \cite{Kilian:2011ka}). Although many QCD/NLO and SM 
background physics topics (\emph{cf.} \emph{e.g.}
\cite{Binoth:2010ra,Butterworth:2010ym,Binoth:2009rv,Greiner:2011mp})
have been tackled using the package, its main focus lies on BSM physics and its collider
phenomenology (\emph{cf.} e.g. \cite{Ohl:2004tn}).  
Specifically models focusing on alternative scenarios have
been incorporated in {\sc Whizard} and independently been tested and
validated~\cite{Ohl:2010zf,Ohl:2008ri,Alboteanu:2008my,Kilian:2006eh,Kilian:2006cj,Beyer:2006hx,Kilian:2004pp},
where specifically also new physics effects from higher-dimensional
operators have been investigated. The new interface to {\sc FeynRules}
described in this paper makes such studies considerably easier. 
\begin{figure*}
\centerline{\includegraphics[width=0.9\textwidth]{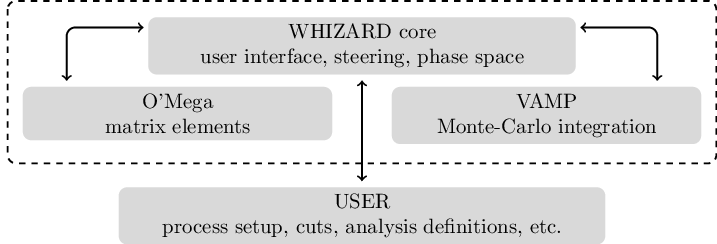}}
\caption{General structure of the {\sc Whizard} package.}
\label{fig-struct-whiz}
\end{figure*}

The program is not designed as a monolithic block, but instead shows a
high degree of modularity including the, in principle, independent and
stand-alone usable matrix-element generator {\sc O'Mega} and the adaptive
multi-channel Monte Carlo integrator {\sc Vamp} as well as several
dedicated sub-packages for lepton collider physics, all of which
are contained in an easy-to-use package. This design is shown schematically in
\figref{fig-struct-whiz}. {\sc Whizard} has been designed such that this
structure is completely transparent to the user.

For providing additional functionality like convolution with parton
distributions, hadronization or support for additional event formats,
{\sc Whizard} can be linked against external packages
like {\sc Lhapdf} \cite{Whalley:2005nh}, {\sc Pythia}
\cite{Sjostrand:2006za} or also
{\sc HepMC} \cite{Dobbs:2001ck}.
\\

\paragraph{\textbf{\textsc{O'Mega}:}}

O'Mega \cite{Moretti:2001zz}
is a generator for tree-level matrix elements which generates a symbolic
representation of the scattering amplitude and translates it into
{\sc Fortran 90} code. As this is a computer-algebraic process
which is rather similar to the workflow
of an optimizing compiler, the {\sc OCaml} language has been chosen
for the implementation of {\sc O'Mega}. 

The algorithm is based on the algebraic paradigm of so-called Directed
Acyclical Graphs (DAGs) which allow a re-usage of all already computed
components. Inspired by the ideas of {\sc Helas} \cite{HELAS}, amplitudes
are constructed by building a
representation of the matrix element as a tree of one-particle
irreducible wave functions (1POWs, Greens functions with all but one
leg amputated) which is then translated into highly optimized {\sc Fortran
90} code to evaluate the matrix element by numerically fusing
1POWs. This approach avoids the repeated and redundant evaluation of
subdiagrams present in any Feynman diagram based approach and brings the
factorial growth in complexity of the latter down to an exponential behavior.
In addition, the optimized structure of the amplitudes is especially
well-suited to cope with big cancellations present in theories with
gauge invariances.

The algebraic {\sc O'Mega} algorithm
bears similarities to the numerical recursive solution
of Schwinger-Dyson equations of motion implemented in the code
{\sc Alpha} \cite{Caravaglios:1996nq}. In this program, however,
the evaluation proceeds fully numerically and no symbolic representation
of the amplitude is constructed.

While there is no design limitation on the spin of the fields or on their
interactions, {\sc O'Mega} currently supports spin
$0$, $\frac{1}{2}$, $1$, $\frac{3}{2}$
and $2$ fields, most dimension three and four couplings between those
(in particular all dimension four gauge interactions) and a small set of higher
dimension operators\footnote{
A future revision of {\sc Whizard/O'Mega}  will allow
specifying the Lorentz structure of the couplings directly in the
model file, thus alleviating the limitation to predefined structures
and permitting a much larger set of couplings. 
}.

Color treatment in {\sc O'Mega} leverages the color flow \cite{Maltoni:2002mq,CFC}
decomposition and currently supports fields transforming either as
singlets, triplets, antitriplets or octets. Only the vertex factor
belonging to a predefined flow
is implemented in the model definition, and the other flows are derived
by {\sc O'Mega} from the $\mathbf{SU}(3)_\text{C}$ representations of the fields.

Models are represented in {\sc O'Mega} as {\sc OCaml} modules which contain symbolic
definitions of the fields and coupling constants, vertex lists, translation
functions which map the fields onto their Lorentz / $\mathbf{SU}(3)_\text{C}$
representations and a number of auxiliary maps which provide textual
representations of the fields and constants for interfacing and code generation
purposes. These modules are compiled and linked with the {\sc O'Mega} framework to
obtain executables which can be called to transform a process definition into
{\sc Fortran} code.
\\

\paragraph{\textbf{\textsc{Vamp}:}}

{\sc Vamp} \cite{Ohl:1998jn}
is a multichannel extension of the {\sc Vegas} \cite{Lepage} algorithm. For all
possible singularities in the integrand, suitable maps and integration channels
are chosen which are then weighted and superimposed to build the phase space
parameterization. Both grids and weights are modified in the adaption phase of
the integration.

The multichannel integration algorithm is implemented as a {\sc Fortran 90}
library with the task of mapping out the integrand and finding suitable
parameterizations being completely delegated to the calling program ({\sc
Whizard} core in this case). This makes the actual {\sc Vamp} library completely
agnostic of the model under consideration.
\\

\paragraph{\textbf{\textsc{Whizard} core:}}

With matrix element generation and Monte Carlo integration being outsourced into
dedicated building blocks, it is the job of the program core
to roll {\sc O'Mega} and {\sc Vamp} transparently into a package which can be
operated without dealing with the components directly.

To this end, the core provides a command-line based user interface
which is used to specify the process, event generation and analysis
setup and to control the runs of the event generator. In addition, it
is responsible for all physics pieces not covered by the other two
components, most importantly phase space generation and beam (parton densities)
setup. 

The major task of the {\sc Whizard} core is
to generate the phase space and map out the integrand, for which it needs
information on the model under consideration. 
In the legacy 1.x branch of {\sc Whizard}, the core is implemented as a set of
{\sc Fortran}
90 modules, {\sc Perl} glue for managing the code generation and Makefiles for
steering the build process. New models have to be added to the build framework
in order to be available to the user.

The recently published 2.x branch of {\sc Whizard} supersedes the old versions and
features a complete rewriting of the core as a single {\sc Fortran 2003} program,
discarding the old {\sc Perl} code. Models are loaded dynamically and searched for in
a configurable search path, allowing the user to add new models without
modifying the {\sc Whizard} package itself. Moreover, the dedicated scripting
language {\sc Sindarin} has been introduced which steers all aspects of the
simulation and which replaces the set of input files used in previous versions.


\section{Implementing new models in Whizard using FeynRules}
\label{sec:WO_marries_FR}

\subsection{General strategy}
In this Section, we describe how to implement generic BSM models into 
{\sc Whizard} in  an efficient way by means of the newly developed 
interface from  {\sc FeynRules} to {\sc Whiz\-ard}.

This new interface brings the implementation of new 
models into {\sc Whizard} in line with the approach introduced in 
Ref.~\cite{Christensen:2009jx}, where its power to develop, implement and 
validate models has been demonstrated in the context of 
{\sc CalcHep/CompHep}, {\sc MadGraph/MadEvent} and {\sc Sherpa}.
This approach is built around {\sc FeynRules}, a {\sc Mathematica}
package where any perturbative and local quantum field theory Lagrangian 
can be implemented and its Feynman rules obtained in an automated way.
The interaction vertices can then be passed, together with all relevant 
information, to various matrix element generators through dedicated 
interfaces.

The new {\sc FeynRules} interface to {\sc Whizard} allows the user to integrate 
any model implemented in {\sc FeynRules} into the {\sc Whizard/O'Mega} 
framework, although some technical restrictions apply and will be explained in
the next section. The interface produces a set of files that contain all the particle and
parameter definitions of the model together with the interaction vertices. 
Together with the model files, a framework is created which allows to
communicate the new models to {\sc Whizard} in a well defined way, after which
step the model can be used exactly like the built-in ones.
This specifically means that the user is not required to
manually modify the code of
{\sc Whizard/O'Mega}, the models created by the interface can be used
directly without any further user intervention.

\subsection{Features and restrictions of the {\sc Whizard} -- {\sc FeynRules} interface}

In this Section we describe in detail the features and restrictions of the new
interface from {\sc FeynRules} to {\sc Whiz\-ard}. After initializing {\sc
FeynRules} and loading a model definition, the code can be invoked in
a similar
fashion as the other interfaces described in Ref.\ \cite{Christensen:2009jx} in
order to transform the Feynman  rules obtained automatically from a Lagrangian
into a set of model files ready  to use with {\sc {\sc Whizard}}.
Both the legacy branch 1.9x as well  as the current 2.x series
of {\sc Whizard} are actively supported by the code generator.
Options that can be passed to the interface are introduced only as needed, while
a detailed explanation of all options can be found in
\appref{app:interface-options}.
\\

\paragraph{\bf Supported fields and vertices:}
All fields currently supported by {\sc FeynRules} (scalars, Dirac and Majorana
fer\-mions, vectors and symmetric tensors) are handled by the interface. 
The set of
accepted operators, the full list of which can be found in
\tabref{tab-operators}, is a subset of all the operators supported by {\sc O'Mega}.
While still limited, this list is sufficient for a large number of BSM models.
In addition, a future version of {\sc
  Whizard/O'Mega} will support the definition of completely general 
Lorentz structures in the model, allowing the interface to
translate all interactions handled by {\sc FeynRules}.
\\

\begin{table*}[!t]
\centerline{\begin{tabular}{|c|c|}
\hline Particle spins & Supported Lorentz structures \\\hline\hline
FFS & \parbox{0.7\textwidth}{\raggedright
   All operators of dimension four are supported.
\strut}\\\hline
FFV & \parbox[t]{0.7\textwidth}{\raggedright
   All operators of dimension four are
   supported.
\strut}\\\hline
SSS & \parbox{0.7\textwidth}{\raggedright
   All dimension three interactions are supported.
\strut}\\\hline
SVV & \parbox[t]{0.7\textwidth}{\raggedright
   Supported operators:\\
   \mbox{}\hspace{5ex}$\begin{aligned}
      \text{dimension 3:} & \quad\mathcal{O}_3 = V_1^\mu V_{2\mu}\phi \mbox{}\\
      \text{dimension 5:} & \quad\mathcal{O}_5 = \phi
         \left(\partial^\mu V_1^\nu - \partial^\nu V_1^\mu\right)
         \left(\partial_\mu V_{2\nu} - \partial_\nu V_{2\mu}\right)
   \end{aligned}$\\
Note that $\mathcal{O}_5$ generates the effective gluon-gluon-Higgs couplings obtained by integrating out heavy quarks.
\strut}\\\hline
SSV & \parbox[t]{0.7\textwidth}{\raggedright
   $\left(\phi_1\partial^\mu\phi_2 - \phi_2\partial^\mu\phi_1\right)V_\mu\;$
   type interactions are supported.
\strut}\\\hline
SSVV & \parbox{0.7\textwidth}{\raggedright
   All dimension four interactions are supported.
\strut}\\\hline
SSSS & \parbox{0.7\textwidth}{\raggedright
   All dimension four interactions are supported.
\strut}\\\hline
VVV & \parbox[t]{0.7\textwidth}{\raggedright
   All parity-conserving dimension four operators are supported, with the
   restriction that non-gauge interactions may be split into several vertices and
   can only be handled if all three fields are mutually different.\strut
\strut}\\\hline
VVVV & \parbox[t]{0.7\textwidth}{\raggedright
   All parity conserving dimension four operators are supported.
\strut}\\\hline
TSS, TVV, TFF & \parbox[t]{0.7\textwidth}{\raggedright
   The three point couplings in the Appendix of Ref.\ \cite{Han:1998sg} are supported.
\strut}\\\hline
\end{tabular}}
\caption{All Lorentz structures currently supported by the {\sc Whizard} -- {\sc
    FeynRules} interface, sorted with respect to the spins of the
  particles. ``S'' stands for scalar, ``F'' for fermion (either
  Majorana or Dirac) and ``V'' for vector.} 
\label{tab-operators}
\end{table*}

\paragraph{\bf Color:}
Color is treated in {\sc O'Mega} in the color flow decomposition,
with the flow structure being implicitly determined from
the representations of the particles present at the vertex. Therefore, the
interface has to strip the color structure from the vertices derived by
{\sc FeynRules} before writing them out to the model files.
While this process is
straightforward for all color structures which correspond only to a single flow
assignment, vertices with several possible flow configurations must be treated
with care in order to avoid mismatches between the flows assigned by {\sc
O'Mega} and those actually encoded in the couplings. To this end, the interface
derives the color flow decomposition from the color structure determined by
{\sc FeynRules} and rejects all vertices which would lead to a wrong flow
assignment by {\sc O'Mega} (these rejections are accompanied by warnings from
the interface).

At the moment, the $\mathbf{SU}(3)_\text{C}$ representations supported by 
both {\sc Whizard} and the interface are singlets ($1$), triplets ($3$), antitriplets 
($\bar{3}$) and octets ($8$).
\tabref{tab:su3struct} shows all combinations of these representations which can
form singlets together with the support status of the respective color
structures in {\sc Whizard} and the interface. Although the supported
color structures do not
comprise all possible singlets, the list is sufficient for a large number of
SM extensions. Furthermore, a future revision of {\sc
Whizard/O'Mega} will allow for explicit color flow assignments, thus
removing most of the current restrictions.

In models generated for the 1.9x branch of {\sc Whizard}, a hardcoded limit exists
on the maximum number $n_\text{cf}$ of external color lines which is related to
the number of external octets $n_8$ and triplets / antitriplets $n_3$ as
\begin{equation*}
n_\text{cf} = n_8 + \frac{n_3}{2} \;\; .
\end{equation*}
This limit is set to $n_\text{cf}=4$ by default and can be changed via the
option \texttt{WOMaxNcf}. No such limit applies to {\sc Whizard} 2.0.0 and higher.
\begin{table*}
\centerline{\begin{tabular}{|c|c|}
\hline $\mathbf{SU}(3)_\text{C}$ representations &
   Support status
\\\hline\hline
\parbox[t]{0.2\textwidth}{
   \centerline{\begin{tabular}[t]{lll}
   $111,\quad$ & $\bar{3}31,\quad$ & $\bar{3}38,$ \\
   $1111,$ & $\bar{3}311,$ & $\bar{3}381$
   \end{tabular}}} &
\parbox[t]{0.7\textwidth}{\raggedright\strut Fully supported by the interface\strut}
\\\hline
$888,\quad 8881$ &
\parbox{0.7\textwidth}{\raggedright\strut Supported only if at least two of the octets
are identical particles.\strut}
\\\hline
$881,\quad 8811$ &
\parbox{0.7\textwidth}{\raggedright\strut Supported in output for {\sc Whizard} 1.96
and higher (including the new 2.x branch) only\footnote{%
Note that in order to use such couplings in the 1.9x branch, the O'Mega option \texttt{2g}
must be added to the process definition in \texttt{whizard.prc}.}.\strut}
\\\hline
$\bar{3}388$ &
\parbox{0.7\textwidth}{\raggedright\strut Supported only if the octets are identical
particles.\strut}
\\\hline
$8888$ &
\parbox{0.7\textwidth}{\raggedright\strut The only supported flow structure is
\begin{equation*}
\parbox{21mm}{\includegraphics{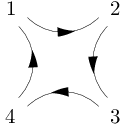}}\cdot\;\Gamma(1,2,3,4)
   \quad+\quad \text{all acyclic permutations}
\end{equation*}
where $\Gamma(1,2,3,4)$ represents the Lorentz structure associated with the
first flow.\strut}
\\\hline
\parbox[t]{0.2\textwidth}{
   \centerline{\begin{tabular}[t]{lll}
   $333,\quad$ & $\bar{3}\bar{3}\bar{3},\quad$ & $3331$\\
   $\bar{3}\bar{3}\bar{3}1,$ & $\bar{3}\bar{3}33$
   \end{tabular}}} &
\parbox[t]{0.7\textwidth}{\raggedright\strut Unsupported.\strut}
\\\hline
\end{tabular}}
\caption{All possible combinations of three or four $\mathbf{SU}(3)_\text{C}$
representations supported by {\sc FeynRules} which can be used to build singlets,
together with the support status of the corresponding color structures in
{\sc Whizard} and the interface.}
\label{tab:su3struct}
\end{table*}
\\

\paragraph{\bf Running $\alpha_S$:}

Starting with version 2.0, {\sc Whizard} supports a running strong coupling. While
this is fully supported by the interface, a choice has to be made which
quantities are to be reevaluated when the strong coupling is evolved. We choose to
implement the running such that by default \texttt{aS},
\texttt{G} (see Ref. \cite{Christensen:2008py} for the nomenclature regarding
the QCD coupling) and any vertex factors depending on them are evolved.
The list of internal parameters that are to be recalculated (together with the
vertex factors depending on them) can be
extended (beyond \texttt{aS} and \texttt{G}) by using
the option \texttt{WORunParameters} when calling the interface.
As older versions of {\sc Whizard} do not evolve the strong coupling, the interface
does not support running with {\sc Whizard} 1.x.
\\

\paragraph{\bf Gauge choices:}
\label{sec:gauge-choices}
The interface supports the unitarity, Feynman and $R_\xi$ gauges. The choice of
gauge must be communicated to the interface via the option \texttt{WOGauge}.
Note that massless gauge bosons are always treated in Feynman gauge.

If the selected gauge is Feynman or $R_\xi$, the interface can
automatically assign the proper masses to the Goldstone bosons. This behavior is
requested by using the \texttt{WOAutoGauge} option. In the $R_\xi$
gauges, the symbol representing the gauge $\xi$ must be communicated to the
interface by using the \texttt{WOGaugeSymbol} option (the symbol is
automatically introduced into the list of external
parameters if \texttt{WOAutoGauge} is
selected at the same time). This feature can be used to automatically extend
models implemented in Feynman gauge to the $R_\xi$ gauges.

Since {\sc Whizard} is a tree-level tool working with helicity amplitudes, the ghost sector is
irrelevant for {\sc Whizard} and can be left out. Ghosts and interactions involving
ghosts are dropped by the interface.

\subsection{Usage}
\label{sec:interface-usage}

\paragraph{{\bf Installation and basic usage:}}
From version 1.6.0 onward, the interface is included into the official
{\sc FeynRules} distribution. In addition, the latest version can be downloaded from
the {\sc Whizard} homepage on {\sc HepForge}. An installer is included
which installs the interface into an existing {\sc FeynRules}
installation (this allows to use the program with {\sc FeynRules}
1.4.x where it is not part of the package). 

Once installed, the interface can be called and used in the same way as the
other interfaces described in Ref.\ \cite{Christensen:2008py}. For example, once the
{\sc FeynRules} environment has been initialized and a model has been loaded, the
command
\begin{quotation}\texttt{
WriteWOOutput[L]
}\end{quotation}
will call \texttt{FeynmanRules} to extract the Feynman rules from the Lagrangian
\texttt{L}, translate them together with the model data and finally write the
files necessary for using the model with {\sc Whizard} to an output directory (the
name of which is inferred from the model name by default). Options can be
added for further control over the translation process (see
\appref{app:interface-options}).
Instead of using a Lagrangian, it is also possible to call the
interface on a vertex list. For example, the following command
\begin{quotation}\texttt{
WriteWOOutput[Input -> list]
}\end{quotation}
will directly translate the vertex list \texttt{list}.  Note that this vertex list must
be given in flavor-expanded form in order for the interface to process it correctly.

The interface also supports the \texttt{WriteWOExtParams} command described in 
Ref.\ \cite{Christensen:2008py} for the other interfaces.  Issuing
\begin{quotation}\texttt{
WriteWOExtParams[filename]
}\end{quotation}
will write a list of all the external parameters to \texttt{filename}.
The format of the output depends on the {\sc Whizard} version it is targeted at: for
1.9x, it is a {\sc Fortran} namelist suitable for use in \texttt{whizard.in}, while a
{\sc Sindarin} script is generated for 2.x. The target version is selected with the
option \texttt{WOWhizardVersion} (which is the only option accepted by the
command).

During execution, the interface will print out a series of messages. It is highly
advised to carefully read through this output as it not only summarizes the settings and
the location of the output files, but also contains information on any skipped
vertices or potential incompatibilities of the model with {\sc Whizard}.\\

\paragraph{{\bf Propagating the output to {\sc Whizard}:}}
After the interface has run successfully and written the model files to the
output directory, the model must be imported into {\sc Whizard}. This step depends on
whether the code has been generated for the legacy 1.9x branch or for 2.x.

In the case of version 1.9x, the model files must be patched into the {\sc Whizard}
tree before the model can be used. For this purpose, a script called
\texttt{inject.sh} is created in the output directory. For most applications, this
step is as easy as issuing (assuming that the output directory is the current
working directory)
\begin{quotation}\texttt{
./inject.sh /path/to/whizard
}\end{quotation}
(replacing the path with the correct path to the {\sc Whizard} installation). After
reconfiguring and recompiling {\sc Whiz\-ard}, the model is available and can  be used
in a similar fashion as the built-in ones. The installer supports several
options for more complex usage scenarios which are described in the
\texttt{INSTALL} created in the output directory.

For the new 2.x branch, the model files are compiled and can then be installed
independently of {\sc Whizard}.
In the simplest scenario, assuming that the output directory
is the current working directory and that the {\sc Whizard} binaries can be found in
the current \texttt{\$\{PATH\}}, the installation is performed by issuing
\begin{quotation}\texttt{
./configure~\&\&~make clean~\&\&~make install
}\end{quotation}
This will compile the model and install it into the directory
\texttt{\$\{HOME\}/.whizard}, making it fully available to {\sc Whizard} without any
further intervention. The build system can be adapted to more complicated cases
through several options to \texttt{configure} which are listed in the
\texttt{INSTALL} file created in the output directory.

\subsection{Validation of the interface}

The output of the interface has been extensively validated for both the
1.9x and 2.x branches. Specifically, the integrated cross sections for all possible
$2\rightarrow 2$ processes in the {\sc FeynRules} SM, the MSSM and the
Three-Site Higgsless Model \cite{Chivukula:2006cg} have
been compared between {\sc Whizard}, {\sc MadGraph} and
{\sc CalcHep} (for all three of which interfaces are included in {\sc
  FeynRules}). In addition, the native versions of these models for
the respective codes have also been included into the
comparison\footnote{ 
A native version of the Three-Site Model is available only for CalcHEP and
{\sc Whizard}.
}. For those codes which support it ({\sc CalcHep} and {\sc Whizard}), different gauges
have been checked as well.
In all comparisons, excellent agreement within the Monte Carlo errors was
achieved. Several tables showcasing the results for selected processes in all three models
can be found in \appref{app:validation-tables}.

\section{Usage examples}
\label{sec:usageex}

In order to illustrate the usage of the interface as detailed above, we will
now show several examples of generating output for {\sc Whizard} 2. The examples
are constructed to show the application of the different options of the
interface and to serve as a starting point for the generation of {\sc Whizard}
versions of other {\sc FeynRules} models.

\subsection{Standard Model}\label{sec:usage SM}

To start off, we will create {\sc Whizard} 2 versions of the Standard Model as implemented in
{\sc FeynRules} for different gauge choices.

\paragraph{{\bf Unitarity Gauge}} 

In order to invoke {\sc FeynRules}, we change to the corresponding directory
and load the program in {\sc Mathematica} via
\\[2ex]\mbox{}\hspace{3ex}\begin{minipage}{0.9\textwidth}
\begin{verbatim}
$FeynRulesPath = 
     SetDirectory["<path-to-FeynRules>"];
<<FeynRules`
\end{verbatim}\end{minipage}
\\[2ex]
The model is loaded by
\\[2ex]\mbox{}\hspace{3ex}\begin{minipage}{0.9\textwidth}
\begin{verbatim}
LoadModel["Models/SM/SM.fr"];
FeynmanGauge = False;
\end{verbatim}\end{minipage}
\\[2ex]
Note that the second line is required to switch the Standard
Model to Unitarity gauge as opposed to Feynman gauge (which is the default).
Generating a {\sc Whizard 2} version of the model is now as easy as doing
\\[2ex]\mbox{}\hspace{3ex}\begin{minipage}{0.9\textwidth}
\begin{verbatim}
WriteWOOutput[LSM];
\end{verbatim}\end{minipage}
\\[2ex]

After invokation, the interface first gives a short summary of the setup
\\[2ex]\mbox{}\hspace{3ex}\begin{minipage}{0.9\textwidth}
\begin{verbatim}
Short model name is "fr_standard_model"
Gauge: Unitarity
Generating code for WHIZARD / O'Mega 
                        version 2.0.3
Maximum number of couplings per FORTRAN 
                           module: 500 
Extensive lorentz structure checks disabled.
\end{verbatim}\end{minipage}
\\[2ex]
Note that, as we have not changed any options, those settings represent the
defaults. The output proceeds with the calculation of the Feynman rules from the
Standard Model Lagrangian \verb?LSM?. After the rules have been derived, the
interface starts generating output and tries to match the vertices to their {\sc
Whizard} / {\sc O'Mega} counterparts.
\\[2ex]\mbox{}\hspace{3ex}\begin{minipage}{0.9\textwidth}
\begin{verbatim}
   10 of 75 vertices processed...
   20 of 75 vertices processed...
   30 of 75 vertices processed...
   40 of 75 vertices processed...
   50 of 75 vertices processed...
   60 of 75 vertices processed...
   70 of 75 vertices processed...
processed a total of 75 vertices, kept 74 
   of them and threw away 1, 1 of which 
   contained ghosts or goldstone bosons.
\end{verbatim}\end{minipage}
\\[2ex]
The last line of the above output is particularily interesting, as it informs us
that everything worked out correctly: the interface was able to match all
vertices, and the only discarded vertex was the QCD ghost interaction.
After the interface has finished running, the model files in the output
directory are ready to use and can be compiled using the procedure described in
the previous section.

\paragraph{{\bf Feynman and $R_\xi$ gauges}}

As the Standard Model as implemented in {\sc FeynRules} also supports Feynman
gauge, we can use the program to generate a Feynman gauge version of the model.
Loading {\sc FeynRules} and the model proceeds as above, with the only
difference being the change
\\[2ex]\mbox{}\hspace{3ex}\begin{minipage}{0.9\textwidth}
\begin{verbatim}
FeynmanGauge = True;
\end{verbatim}\end{minipage}
\\[2ex]
In order to inform the interface about the modified gauge, we have to add the
option \verb?WOGauge?
\\[2ex]\mbox{}\hspace{3ex}\begin{minipage}{0.9\textwidth}
\begin{verbatim}
WriteWOOutput[LSM, WOGauge -> WOFeynman];
\end{verbatim}\end{minipage}
\\[2ex]
The modified gauge is reflected in the output of the interface
\\[2ex]\mbox{}\hspace{3ex}\begin{minipage}{0.9\textwidth}
\begin{verbatim}
Short model name is "fr_standard_model"
Gauge: Feynman
Generating code for WHIZARD / O'Mega 
                        version 2.0.3
Maximum number of couplings per FORTRAN 
                           module: 500
Extensive lorentz structure checks disabled.
\end{verbatim}\end{minipage}
\\[2ex]
The summary of the vertex identification now takes the following form
\\[2ex]\mbox{}\hspace{3ex}\begin{minipage}{0.9\textwidth}
\begin{verbatim}
processed a total of 163 vertices, kept 139 
   of them and threw away 24, 24 of which 
   contained ghosts.
\end{verbatim}\end{minipage}
\\[2ex]
Again, this line tells us that there were no problems --- the only discarded
interactions involved the ghost sector which is irrelevant for {\sc Whizard}.

As {\sc Whizard} does not depend on the ghost sector, the only difference
between the different gauges from the perspective of the interface are the gauge
boson propagators and the Goldstone boson masses. Therefore, the interface can
automatically convert a model in Feynman gauge to a model in $R_\xi$ gauge. To
this end, the call to the interface must be changed to
\\[2ex]\mbox{}\hspace{3ex}\begin{minipage}{0.9\textwidth}
\begin{verbatim}
WriteWOOutput[LSM, WOGauge -> WORxi, 
               WOAutoGauge -> True];
\end{verbatim}\end{minipage}
\\[2ex]
The \verb?WOAutoGauge? argument instructs the interface to automatically
\begin{enumerate}
\item Introduce a symbol for the gauge parameter $\xi$ into the
list of external parameters
\item Generate the Goldstone boson masses from those of the associated gauge
bosons (ignoring the values provided by {\sc FeynRules})
\end{enumerate}
The modified setup is again reflected in the interface output
\\[2ex]\mbox{}\hspace{3ex}\begin{minipage}{0.9\textwidth}
\begin{verbatim}
Short model name is "fr_standard_model"
Gauge: Rxi
Gauge symbol: "Rxi"
Generating code for WHIZARD / O'Mega 
                       version 2.0.3
Maximum number of couplings per FORTRAN 
                         module: 500
Extensive lorentz structure checks disabled.
\end{verbatim}\end{minipage}
\\[2ex]
Note the default choice \verb?Rxi? for the name of the $\xi$ parameter --- this
can be modified via the option \verb?WOGaugeParameter?.

While the \verb?WOAutoGauge? feature allows to generate $R_\xi$ gauged models
from models implemented in Feynman gauge, it is of course also possible to use
models genuinely implemented in $R_\xi$ gauge by setting this parameter to
\verb?False?. Also, note that the choice of gauge only affects the propagators
of massive fields. Massless gauge bosons are always treated in Feynman
gauge.

\paragraph{{\bf Compilation and usage}}

In order to compile and use the freshly generated model files, change to the
output directory which can be determined from the interface output (in this
example, it is \verb?fr_standard_model-WO?). Assuming that {\sc Whizard} is
available in the binary search path, compilation and installation proceeds as
described in the previous Section by issuing (from the shell)
\\[2ex]\mbox{}\hspace{3ex}\begin{minipage}{0.9\textwidth}
\begin{verbatim}
./configure && make && make install
\end{verbatim}\end{minipage}
\\[2ex]
The model is now ready and can be used similarly to the builtin {\sc
Whizard} models. For example, a minimal {\sc Whizard} input file for calculating
the $e^+e^- \longrightarrow W^+W^-$ scattering cross section in the freshly
generated model would look like
\\[2ex]\mbox{}\hspace{3ex}\begin{minipage}{0.9\textwidth}
\begin{verbatim}
model = fr_standard_model
process test = "e+", "e-" -> "W+", "W-"
sqrts = 500 GeV
integrate (test)
\end{verbatim}\end{minipage}
\\[2ex]

\subsection{Minimal Supersymmetric Standard Model}
In this Section, we illustrate the usage of the interface between {\sc
FeynRules} and {\sc Whizard} in the context of the MSSM. For the conventions
on the names of the particles and the relative signs entering the different
terms of the Lagrangian, we refer to Refs.\ \cite{Christensen:2009jx,Duhr:2011se}. All the
parameters of the model are then ordered in Les Houches blocks and counters following
the SUSY Les Houches Accord (SLHA)
\cite{Skands:2003cj,Allanach:2008qq}. Let us emphasize that implementing
any model in {\sc FeynRules} only requires a SLHA-like structure, whilst the
choice of the SLHA conventions for the MSSM is only due to the author of the
model implementation.

The neutralino sector deserves special
attention. After diagonalization of the mass matrix expresssed in terms
of the gaugino and higgsino eigenstates, the resulting mass eigenvalues may be
either negative or positive. In this case, two procedures can be followed. 
Either the masses are rendered
positive and the associated mixing matrix gets purely imaginary entries or the
masses are kept signed, the mixing matrix in this case being real. 
According to the SLHA agreement, the second option is adopted.
For a specific eigenvalue, the phase is absorbed into the
definition of the relevant eigenvector, rendering the mass negative. However,
{\sc Whizard} does not officially support negative
masses. For the case of the interface to {\sc FeynRules} this means,
that one must be careful using a SLHA file with explicit factors of
the complex unity in the mixing matrix, and on the other hand, 
real and positive masses for the neutralinos. For the hard-coded SUSY
models, this is completely handled internally, and the SUSY
implementations in WHIZARD have been extensively
tested~\cite{Reuter:2009ex,Kalinowski:2008fk,Robens:2008sa,Hagiwara:2005wg,AguilarSaavedra:2005pw,Ohl:2002jp}.    
Especially Ref.~\cite{Hagiwara:2005wg} discusses the details of the
neutralino (and chargino) mixing matrix. 

After having downloaded the model
from the {\sc FeynRules} website, we store it in a new directory, labelled 
\verb"MSSM", of the model library of the local installation of {\sc FeynRules}.
The model can then be loaded in {\sc Mathematica} as
\\[2ex]\mbox{}\hspace{3ex}\begin{minipage}{0.9\textwidth}
\begin{verbatim}
$FeynRulesPath = 
        SetDirectory["<path-to-FeynRules>"];
<<FeynRules`
LoadModel["Models/MSSM/MSSM.fr"];
FeynmanGauge = False;
\end{verbatim}\end{minipage}
\\[2ex]
As it can be seen from the last command line, we have adopted the unitarity
gauge. Other choices, such as Feynman gauge, are also possible, and we refer 
to Section \ref{sec:usage SM} for using the interface with other gauge choice. 

The number of vertices associated to supersymmetric Lagrangians is in general
very large (several thousands). For such models with many interactions,
it is recommended to first extract all the Feynman rules of the theory before
calling the interface between {\sc FeynRules} and {\sc Whizard}. 
The reason is related to the efficiency of the interface  which takes a lot of
time in the extraction of the interaction vertices. In the case
one wishes to study the phenomenology of several benchmark scenarios, this
procedure, which is illustrated below,
allows to use the interface in the most optimal way. The Feynman rules
are derived from the Lagrangian once and for all and then reused by the
interface for each set of {\sc Whizard} model files to be produced, considerably 
speeding up the generation of multiple model files issued from a single
Lagrangian.
In addition, the
scalar potential of supersymmetric theories contains a large set of four scalar
interactions, in general irrelevant for collider phenomenology. These vertices
can be neglected with the help of the \verb"Exclude4Scalars->True" option of
both the commands \verb"FeynmanRules" and \verb"WriteWOOutput".  The Feynman
rules of the MSSM are then computed by issuing in the {\sc Mathematica} notebook,
\\[2ex]\mbox{}\hspace{3ex}\begin{minipage}{0.9\textwidth}
\begin{verbatim}
rules = FeynmanRules[lag, 
   Exclude4Scalars->True, FlavorExpand->True];
\end{verbatim}\end{minipage}
\\[2ex]
where \verb'lag' is the variable containing the Lagrangian.

By default, all the parameters of the model are set to the value of $1$. A
complete 
parameter file must therefore be loaded. We choose the
SPS1a~\footnote{Though we are aware that this specific point is
  excluded by LHC data, we use it here purely for demonstration
  purposes, as the validation of the MSSM within the interface has
  been done with it.} benchmark point
\cite{Allanach:2002nj}, where we take care of setting all the masses of the
neutralinos to positive values and adapt the associated mixing
matrix accordingly. This parameter file, named \verb"sps1a_wo.dat",
can be downloaded from the {\sc FeynRules} website, and loaded into {\sc
FeynRules} as
\\[2ex]\mbox{}\hspace{3ex}\begin{minipage}{0.9\textwidth}
\begin{verbatim}
ReadLHAFile[Input -> "sps1a_wo.dat"];
\end{verbatim}\end{minipage}
\\[2ex]
We note that this command does not reduce the size of the model output by removing
vertices with vanishing couplings.  However, if desired, this task could be done with the 
\texttt{LoadRestriction} command (see Ref.\ \cite{Fuks:2012im} for details).

The vertices are
now ready to be exported to {\sc Whizard} with the help of the command
\\[2ex]\mbox{}\hspace{3ex}\begin{minipage}{0.9\textwidth}
\begin{verbatim}
WriteWOOutput[Input -> rules];
\end{verbatim}\end{minipage}
\\[2ex]
For compilation and usage of the produced model, we refer to Section
\ref{sec:usage SM}. Finally, let us note that the numerical values of the
parameters of the model can be modified directly at the level of the {\sc
Whizard} program,
without having to generate a second time the {\sc Whizard} model files from {\sc
FeynRules}. To this end, a {\sc Sindarin} script is created by {\sc
FeynRules} with the help of the instruction
\\[2ex]\mbox{}\hspace{3ex}\begin{minipage}{0.9\textwidth}
\begin{verbatim}
WriteWOExtParams["parameters.sin"];
\end{verbatim}\end{minipage}
\\[2ex]
and can be further modified according to the needs of the user when employing the
{\sc Whizard} program.

\subsection{Three-Site Higgsless Model}
The Three-Site  or Minimal Higgsless Model (MHM) \cite{Chivukula:2006cg} is a
minimal deconstructed Higgsless model which contains only the first resonance in
the tower of Kaluza-Klein modes of a Higgsless extra-dimensional
model.    It is a non-renormalizable, effective theory whose
gauge group is an extension of the SM with an extra $SU(2)$ gauge
group.  The breaking of the extended electroweak gauge symmetry is accomplished
by a set of nonlinear sigma fields which represent the effects of
physics at a higher scale and make the theory nonrenormalizable.  The physical
vector boson spectrum contains the usual photon, $W^\pm$ and $Z$ bosons as well
as a $W'^\pm$ and $Z'$ boson.  Additionally, a new set of heavy fermions are
introduced to accompany the new gauge group ``site'' which mix to form the
physical eigenstates.  This mixing is controlled by the small mixing parameter
$\epsilon_L$ which is adjusted to satisfy constraints from precision
observables, such as the S parameter \cite{Chivukula:2005xm}.  

The MHM has been implemented into {\sc LanHEP} \cite{He:2007ge}, {\sc FeynRules}
\cite{Christensen:2009jx} and independently into {\sc Whizard} \cite{Speckner:2010zi},
and the collider phenomenology has been studied by making use of these
implementations \cite{He:2007ge,Ohl:2010zf,Speckner:2010zi}.
Furthermore, the independent
implementations in {\sc FeynRules} and directly into {\sc Whizard} have been compared and
found to agree (see Appendix \ref{app:validation-tables}).  In this Section, we
describe the use of the {\sc Whizard} interface to generate {\sc Whizard} model files for
the MHM.  This model has been implemented in Feynman gauge as well as unitarity
gauge and contains the variable \verb|FeynmanGauge| which can be set to
\verb|True|  or \verb|False|.  When set to \verb|True|, the option
\verb|WOGauge-> WOFeynman| must be used, as described in Appendix
\ref{app:interface-options}.   $R_\xi$ gauge can also be accomplished with this
model by use of the options \verb|WOGauge -> WORxi| and
\verb?WOAutoGauge -> True?.  

Since this model makes use of the nonlinear sigma field
\begin{equation}
\Sigma = 1 + i\pi - \frac{1}{2}\pi^2+\cdots
\end{equation}
many higher dimensional operators are included in the model which are not
supported by {\sc Whizard}.  Although {\sc Whizard} can reject these vertices
and print a warning message to the user, it is preferable to remove the vertices
with the option \verb|MaxCanonicalDimension->4| which is passed to
\verb|FeynmanRules| and restricts the Feynman rules to those of
dimension four and smaller\footnote{We note, at this stage, that {\tt
    MaxCanonicalDimension} is an option of the {\tt FeynmanRules}
  function rather than of the interface, itself. In fact, the
  interface accepts all the options of {\tt FeynmanRules} and simply
  passes them on to the latter.}. 

Since the use of different gauges was already illustrated in Section
\ref{sec:usage SM}, we will simply make use of this interface to
output the model in Feynman gauge.  We load {\sc FeynRules} with 
\\[2ex]\mbox{}\hspace{3ex}\begin{minipage}{0.9\textwidth}
\begin{verbatim}
$FeynRulesPath = 
     SetDirectory["<path-to-FeynRules>"];
<<FeynRules`
\end{verbatim}\end{minipage}
\\[2ex]
where \verb|<path-to-FeynRules>| is the path to the {\sc FeynRules} files.  The MHM is loaded by
\\[2ex]\mbox{}\hspace{3ex}\begin{minipage}{0.9\textwidth}
\begin{verbatim}
SetDirectory["<path-to-MHM>"];
LoadModel["3-Site-particles.fr",
   "3-Site-parameters.fr",
   "3-Site-lagrangian.fr"];
FeynmanGauge = True;
\end{verbatim}\end{minipage}
\\[2ex]
where \verb|<path-to-MHM>| is the path to the directory where the MHM model files are stored and where the output of the {\sc Whizard} interface will be written.   The {\sc Whizard} interface is initiated with the command
\\[2ex]\mbox{}\hspace{3ex}\begin{minipage}{0.9\textwidth}
\begin{verbatim}
WriteWOOutput[LGauge, LGold, LGhost, LFermion, 
   LGoldLeptons, LGoldQuarks,
   MaxCanonicalDimension->4, 
   WOGauge->WOFeynman, WOModelName->"fr_mhm"];
\end{verbatim}\end{minipage}
\\[2ex]
where we have also made use of the option \verb|WOModelName| to change
the name of the model as seen by {\sc Whizard} (see Appendix \ref{app:interface-options}).  As in the SM (see Section \ref{sec:usage SM}), the interface begins by writing a short informational message:
\\[2ex]\mbox{}\hspace{3ex}\begin{minipage}{0.9\textwidth}
\begin{verbatim}
Short model name is "fr_mhm"
Gauge: Feynman
Generating code for WHIZARD / O'Mega 
                       version 2.0.3
Automagically assigning Goldstone 
                        boson masses...
Maximum number of couplings per FORTRAN 
                        module: 500
Extensive lorentz structure checks disabled.
\end{verbatim}\end{minipage}
\\[2ex]
After calculating the Feynman rules and processing the vertices, the interface gives the summary:
\\[2ex]\mbox{}\hspace{3ex}\begin{minipage}{0.9\textwidth}
\begin{verbatim}
processed a total of 922 vertices, kept 633
  of them and threw away 289, 289 of which 
  contained ghosts.
\end{verbatim}\end{minipage}
\\[2ex]
showing that no vertices were missed.  The files are stored in the
directory \verb|fr_mhm| and are ready to be installed and used with {\sc Whizard}.

\section{Conclusions}
\label{sec:conclusions}

In this paper, we presented the interface between {\sc FeynRules}, a program 
to automatically generate Feynman rules from a generic input Lagrangian, to 
the multi-particle event generator {\sc Whizard}. The interface can be used 
both for the legacy version {\sc Whizard} 1.xx as well as the actual release 
line {\sc Whizard} 2.x and is included in {\sc FeynRules} from version 1.6 
on. There has been a quite exhaustive validation of the interface both with 
{\sc FeynRules} models used with other Monte Carlo generators as well as with 
hard-coded implementations of models. We tried to keep the description of the 
interface rather self-contained, but recommend the user to consult both {\sc 
Whizard} and {\sc FeynRules} manuals when performing a calculation with the 
interface. The paper gives a complete overview of the supported particle 
types as well as the Lorentz and color structures of interaction vertices. We 
listed the options and flags which steer the compilation of physics models 
and the transferring of the Feynman rules into the generator. 
In order to give examples of the workflow, we applied it to the three
models used in our validation procedure. From the steps given in
the paper to go from the Lagrangian formulation of that model through
FeynRules to a simulation within Whizard, the user can easily apply the
interface to his or her own preferred model.
The 
presented {\sc FeynRules}/{\sc Whizard} interface hopefully facilitates 
phenomenological studies at present and future collider
experiments. A prime example of using the combination between {\sc
  FeynRules} and an event generator like {\sc Whizard} is to
investigate models whose particle content depend on the point in
parameter space, which is however a separate project and will be
presented elsewhere~\cite{pheno}.


\section*{Acknowledgments}

CS has been supported by the Deutsche Forschungsgemeinschaft through
the Research Training Groups GRK 1147
\textit{Theoretical Astrophysics and Particle Physics} and  GRK 1102
\textit{Physics of Hadron Accelerators}. JRR has been partially supported
by the Ministery of Science and Culture (MWK) of the German state
Baden-W\"urttemberg. BF acknowledges support by the Theory-LHC 
France-initiative of the CNRS/IN2P3. NDC was supported by the US
National Science Foundation under grants PHY-0354226 and PHY-0705682
and by the PITTsburgh Particle physics, Astrophysics and Cosmology Center.
CS would like to thank the Michigan State University and
especially S. Chivukula and L. Simmons for their hospitality and T. Ohl for
encouraging him to start this project.


\begin{appendix}


\section{Options of the {\textsc{Whizard}} interface}
\label{app:interface-options}

In the following we present a comprehensive list of all the options accepted by
\texttt{WriteWOOutput}. Additionally, we note that all options of {\tt FeynmanRules} are accepted by \texttt{WriteWOOutput}, which passes them on to {\tt FeynmanRules}.
\begin{description}
\item[\texttt{Input}]\mbox{}\\
An optional vertex list to use instead of a Lagrangian (which can then be
omitted).
\item[\texttt{WOWhizardVersion}]\mbox{}\\
Select the {\sc Whizard} version for which code is to be generated.
The currently available choices are summarized in \tabref{tab-wowhizardversion}.
\begin{table}
\centerline{\begin{tabular}{|l|l|}
\hline \texttt{WOWhizardVersion} & {\sc Whizard} versions supported \\\hline\hline
\texttt{"1.92"} & 1.92 \\\hline
\texttt{"1.93"} & 1.93 -- 1.95 \\\hline
\texttt{"1.96"} & 1.96+ \\\hline
\texttt{"2.0"} & 2.0.0 -- 2.0.2 \\\hline
\texttt{"2.0.3"} (default) & 2.0.3+ \\\hline
\end{tabular}}
\caption{Currently available choices for the \texttt{WOWhizardVersion} option,
together with the respective {\sc Whizard} versions supported by them.}
\label{tab-wowhizardversion}
\end{table}
This list  will expand as the program evolves. To get a summary
of all choices available in a particular version of the interface, use
the command
\texttt{?WOWhizardVersion}.
\item[\texttt{WOModelName}]\mbox{}\\
The name under which the model will be known to {\sc Whizard}. Please note that models
for versions 1.9x must start with ``\texttt{fr\_}'' if they are to be picked up
by {\sc Whizard} automatically. The default is determined from the {\sc
FeynRules} model
name.
\item[\texttt{Output}]\mbox{}\\
The name of the output directory. The default is determined from the {\sc
FeynRules}
model name.
\item[\texttt{WOGauge}]\mbox{}\\
Gauge choice (\emph{cf.} \secref{sec:gauge-choices}).
Possible values are:\\ \texttt{WOFeynman}, \texttt{WORxi}
and \texttt{WOUnitarity} (default).
\item[\texttt{WOGaugeParameter}]\mbox{}\\
The external or internal parameter representing the gauge $\xi$ in
the $R_\xi$ gauges (\emph{cf.} \secref{sec:gauge-choices}). Default: \texttt{Rxi}
\item[\texttt{WOAutoGauge}]\mbox{}\\
Automatically assign the Goldstone boson masses in the Feynman and $R_\xi$
gauges and automatically append the symbol for $\xi$ to the parameter list in
the $R_\xi$ gauges. Default: \texttt{False}
\item[\texttt{WOMaxNcf}]\mbox{}\\
Maximum number of color flows provided for in the output. Essentially,
this is $n_8 - \frac{1}{2}n_3$ where $n_8$ is the maximum number of external
color octets and $n_3$ is the maximum number of external triplets and
antitriplets. Note that this only affects {\sc Whizard} 1.9x; there is no limit for
2.0.0+. Default: \texttt{4}
\item[\texttt{WORunParameters}]\mbox{}\\
The list of all internal parameters which will be recalculated if $\alpha_S$ is
evolved (see above). This only affects {\sc Whizard} 2.0.0+. Default:
\mbox{\texttt{\{aS, G\}}}
\item[\texttt{WOFast}]\mbox{}\\
If the interface drops vertices which are supported, this option can be
set to \texttt{False} to enable some more time consuming checks which might aid
the identification. Default: \texttt{True}
\item[\texttt{WOMaxCouplingsPerFile}]\mbox{}\\
The maximum number of couplings that are written to a single {\sc Fortran} file. If
compilation takes too long or fails, this can be lowered. Default: \texttt{500}
\item[\texttt{WOVerbose}]\mbox{}\\
Enable verbose output and in particular more extensive information on any
skipped vertices. Default: \texttt{False}
\end{description}

\section{Validation Tables}
\label{app:validation-tables}
In this Appendix we summarize the validation of the interface from
{\sc FeynRules} to {\sc Whizard} 
which was performed
by calculating the integrated cross sections 
related to all possible two-to-two partonic processes in the SM, the MSSM and the 
Minimal Higgsless Model. We evaluated this quantity both with the 
{\sc FeynRules}-generated model files and the built-in (``stock'') implementations of the 
various matrix-element generators, if existing, in the framework of {\sc CalcHep}, 
{\sc MadGraph}/{\sc MadEvent} and {\sc Whizard}/{\sc O'mega}. Where supported by
both model and generator, we also checked both Feynman and unitarity gauge.
We ran the codes such that the Monte Carlo uncertainty was pushed below one percent,
and excellent agreement at this level was achieved in all cases. In 
Tab.~\ref{tab:SM-xs} and \ref{tab:MHM-xs} we present a selection of the results we 
obtained. The full list of results is available from the {\sc FeynRules} website.

In order to  avoid collinear divergences, we imposed a cut on the $p_T$ of each 
final state particle,
\begin{equation}\label{eq_app:ptcut}
p_T\ge {\sqrt{s}\over 4}\,.
\end{equation}
For the MSSM, the parameter set used corresponds to the benchmark point
SPS1a~\cite{Allanach:2002nj} (with the strong coupling set to
$\alpha_s(M_z) = 0.1172$),
whereas for the SM and MHM the values of the input parameters are 
summarized in Tab.~\ref{tab:SM-params} and \ref{tab:MHM-params}. The strong
coupling was evolved to $\sqrt{s}$ for each process. All particle
widths were set to zero.

\begin{table}[!h]
\begin{center}
\begin{tabular}{|r|c |l| }
\hline
Parameter & Symbol & Value\\
\hline\hline
Inverse of the & & \\
electromagnetic coupling  & $\alpha_{EW}^{-1}(M_Z)$ & 127.9\\
Strong coupling  & $\alpha_{s}(M_Z)$ & 0.1172\\
Fermi constant & $G_F$ & \unit[1.16639e-5]{GeV$^{-2}$}\\
$Z$ pole mass & $M_Z$ & \unit[91.188]{GeV} \\
$c$ quark mass & $m_c$ & 0 \\
$b$ quark mass & $m_b$ & \unit[4.7]{GeV}\\
$t$ quark mass & $m_t$ & \unit[174.3]{GeV}\\
$\tau$ lepton mass & $m_\tau$ & \unit[1.777]{GeV}\\
Higgs mass & $m_H$ & \unit[120]{GeV} \\
Cabibbo angle & $\theta_c$ & 0.227736\\
\hline
\end{tabular}
\caption{\label{tab:SM-params} Input parameters for the SM.}
\end{center}
\end{table}

\begin{table}[!h]
\begin{center}
\begin{tabular}{|r|c| l| }
\hline
Parameter & Symbol & Value\\
\hline\hline
Inverse of the & & \\
electromagnetic coupling  & $\alpha_{EW}^{-1}(M_Z)$ & 127.9\\
Strong coupling  & $\alpha_{s}(M_Z)$ & 0.1172\\
$Z$ mass & $M_Z$ & \unit[91.1876]{GeV} \\
$W$ mass & $M_W$ & \unit[80.398]{GeV} \\
$W'$ mass & $M_{W'}$ & \unit[500]{GeV} \\
Heavy fermion mass & $M_F$ & \unit[4]{TeV}\\
$\mu$ lepton mass & $m_\mu$ & \unit[0.1057]{GeV}\\
$\tau$ lepton mass & $m_\tau$ & \unit[1.777]{GeV} \\
$s$ quark mass & $m_s$ & \unit[0.104]{GeV} \\
$c$ quark mass & $m_c$ & \unit[1.27]{GeV} \\
$b$ quark mass & $m_b$ & \unit[4.2]{GeV} \\
$t$ quark mass & $m_t$ & \unit[171.2]{GeV} \\
\hline
\end{tabular}
\caption{\label{tab:MHM-params} Input parameters for the MHM.}
\end{center}
\end{table}

\begin{table*}[!h]
\centerline{\begin{tabular}{|lcl|l|lllllll|lll|}
\hline & & & & \multicolumn{7}{|c|}{\sc FeynRules} & \multicolumn{3}{|c|}{Stock}\\
\multicolumn{3}{|c|}{Process} & \multicolumn{1}{|c|}{$\sqrt{s}$}
 & \multicolumn{1}{|c}{CH} & \multicolumn{1}{c}{CH} & \multicolumn{1}{c}{MG}
 & \multicolumn{1}{c}{WO1} & \multicolumn{1}{c}{WO1} & \multicolumn{1}{c}{WO2}
 & \multicolumn{1}{c|}{WO2} & \multicolumn{1}{|c}{CH} & \multicolumn{1}{c}{MG}
 & \multicolumn{1}{c|}{WO2}\\
& & & \multicolumn{1}{c}{\scriptsize[GeV]} & \multicolumn{1}{|c}{F}
 & \multicolumn{1}{c}{U} & \multicolumn{1}{c}{U} & \multicolumn{1}{c}{F}
 & \multicolumn{1}{c}{U} & \multicolumn{1}{c}{F} & \multicolumn{1}{c|}{U}
 & \multicolumn{1}{|c}{U} & \multicolumn{1}{c}{U} & \multicolumn{1}{c|}{U}\\
\hline\hline
$\scriptstyle W^+\;W^+$ & $\scriptstyle\rightarrow$ & $\scriptstyle W^+\;W^+$ &
 $\scriptstyle 1277$ 
 & {\scriptsize 25.7} & {\scriptsize 25.7} & {\scriptsize 25.7} & {\scriptsize 25.7}
 & {\scriptsize 25.7} & {\scriptsize 25.7} & {\scriptsize 25.7} & {\scriptsize 25.7}
 & {\scriptsize 25.7} & {\scriptsize 25.7}
\\
$\scriptstyle g\;g$ & $\scriptstyle\rightarrow$ & $\scriptstyle g\;g$ &
 $\scriptstyle 200$ 
 & {\scriptsize 1.88e04} & {\scriptsize 1.88e04} & {\scriptsize 1.89e04}
 & {\scriptsize 1.88e04} & {\scriptsize 1.88e04} & {\scriptsize 1.88e04}
 & {\scriptsize 1.88e04} & {\scriptsize 1.88e04} & {\scriptsize 1.88e04}
 & {\scriptsize 1.88e04}
\\
$\scriptstyle Z\;Z$ & $\scriptstyle\rightarrow$ & $\scriptstyle W^+\;W^-$ &
 $\scriptstyle 1368$ 
 & {\scriptsize 26.2} & {\scriptsize 26.2} & {\scriptsize 26.2} & {\scriptsize 26.2}
 & {\scriptsize 26.2} & {\scriptsize 26.2} & {\scriptsize 26.2} & {\scriptsize 26.2}
 & {\scriptsize 26.2} & {\scriptsize 26.2}
\\
$\scriptstyle \gamma\;\gamma$ & $\scriptstyle\rightarrow$ & $\scriptstyle W^+\;W^-$ &
 $\scriptstyle 639$ 
 & {\scriptsize 16.3} & {\scriptsize 16.3} & {\scriptsize 16.3} & {\scriptsize 16.3}
 & {\scriptsize 16.3} & {\scriptsize 16.3} & {\scriptsize 16.3} & {\scriptsize 16.3}
 & {\scriptsize 16.3} & {\scriptsize 16.3}
\\
$\scriptstyle Z\;Z$ & $\scriptstyle\rightarrow$ & $\scriptstyle Z\;Z$ &
 $\scriptstyle 1459$ 
 & {\scriptsize 0.245} & {\scriptsize 0.245} & {\scriptsize 0.245} & {\scriptsize 0.245}
 & {\scriptsize 0.245} & {\scriptsize 0.245} & {\scriptsize 0.245} & {\scriptsize 0.245}
 & {\scriptsize 0.245} & {\scriptsize 0.245}
\\
$\scriptstyle \gamma\;Z$ & $\scriptstyle\rightarrow$ & $\scriptstyle W^+\;W^-$ &
 $\scriptstyle 1003$ 
 & {\scriptsize 19.3} & {\scriptsize 19.3} & {\scriptsize 19.3} & {\scriptsize 19.3}
 & {\scriptsize 19.3} & {\scriptsize 19.3} & {\scriptsize 19.3} & {\scriptsize 19.3}
 & {\scriptsize 19.3} & {\scriptsize 19.3}
\\\hline
$\scriptstyle s\;\bar{s}$ & $\scriptstyle\rightarrow$ & $\scriptstyle Z\;Z$ &
 $\scriptstyle 730$ 
 & {\scriptsize 0.0838} & {\scriptsize 0.0838} & {\scriptsize 0.0839}
 & {\scriptsize 0.0838} & {\scriptsize 0.0838} & {\scriptsize 0.0838}
 & {\scriptsize 0.0838} & {\scriptsize 0.0838} & {\scriptsize 0.0839}
 & {\scriptsize 0.0838}
\\
$\scriptstyle \mu^-\;\mu^+$ & $\scriptstyle\rightarrow$ & $\scriptstyle \gamma\;Z$ &
 $\scriptstyle 366$ 
 & {\scriptsize 1.46} & {\scriptsize 1.46} & {\scriptsize 1.46} & {\scriptsize 1.46}
 & {\scriptsize 1.46} & {\scriptsize 1.46} & {\scriptsize 1.46} & {\scriptsize 1.46}
 & {\scriptsize 1.46} & {\scriptsize 1.46}
\\
$\scriptstyle s\;\bar{s}$ & $\scriptstyle\rightarrow$ & $\scriptstyle \gamma\;\gamma$ &
 $\scriptstyle 200$ 
 & {\scriptsize 0.0272} & {\scriptsize 0.0272} & {\scriptsize 0.0272}
 & {\scriptsize 0.0272} & {\scriptsize 0.0272} & {\scriptsize 0.0272}
 & {\scriptsize 0.0272} & {\scriptsize 0.0272} & {\scriptsize 0.0272}
 & {\scriptsize 0.0272}
\\
$\scriptstyle u\;\bar{d}$ & $\scriptstyle\rightarrow$ & $\scriptstyle Z\;W^+$ &
 $\scriptstyle 684$ 
 & {\scriptsize 0.146} & {\scriptsize 0.146} & {\scriptsize 0.146} & {\scriptsize 0.146}
 & {\scriptsize 0.146} & {\scriptsize 0.146} & {\scriptsize 0.146} & {\scriptsize 0.146}
 & {\scriptsize 0.146} & {\scriptsize 0.146}
\\
$\scriptstyle c\;\bar{c}$ & $\scriptstyle\rightarrow$ & $\scriptstyle g\;g$ &
 $\scriptstyle 200$ 
 & {\scriptsize 462} & {\scriptsize 462} & {\scriptsize 462} & {\scriptsize 462}
 & {\scriptsize 462} & {\scriptsize 462} & {\scriptsize 462} & {\scriptsize 462}
 & {\scriptsize 462} & {\scriptsize 462}
\\
$\scriptstyle \nu_e\;e^+$ & $\scriptstyle\rightarrow$ & $\scriptstyle \gamma\;W^+$ &
 $\scriptstyle 319$ 
 & {\scriptsize 1.98} & {\scriptsize 1.98} & {\scriptsize 1.98} & {\scriptsize 1.98}
 & {\scriptsize 1.98} & {\scriptsize 1.98} & {\scriptsize 1.99} & {\scriptsize 1.98}
 & {\scriptsize 1.98} & {\scriptsize 1.98}
\\
$\scriptstyle d\;\bar{d}$ & $\scriptstyle\rightarrow$ & $\scriptstyle \gamma\;Z$ &
 $\scriptstyle 365$ 
 & {\scriptsize 0.0798} & {\scriptsize 0.0798} & {\scriptsize 0.0799}
 & {\scriptsize 0.0798} & {\scriptsize 0.0798} & {\scriptsize 0.0798}
 & {\scriptsize 0.0798} & {\scriptsize 0.0798} & {\scriptsize 0.0798}
 & {\scriptsize 0.0798}
\\
$\scriptstyle b\;\bar{b}$ & $\scriptstyle\rightarrow$ & $\scriptstyle Z\;Z$ &
 $\scriptstyle 767$ 
 & {\scriptsize 0.0760} & {\scriptsize 0.0760} & {\scriptsize 0.0758}
 & {\scriptsize 0.0760} & {\scriptsize 0.0760} & {\scriptsize 0.0760}
 & {\scriptsize 0.0760} & {\scriptsize 0.0760} & {\scriptsize 0.0758}
 & {\scriptsize 0.0760}
\\
$\scriptstyle c\;\bar{s}$ & $\scriptstyle\rightarrow$ & $\scriptstyle \gamma\;W^+$ &
 $\scriptstyle 325$ 
 & {\scriptsize 0.208} & {\scriptsize 0.208} & {\scriptsize 0.208} & {\scriptsize 0.207}
 & {\scriptsize 0.208} & {\scriptsize 0.208} & {\scriptsize 0.208} & {\scriptsize 0.208}
 & {\scriptsize 0.208} & {\scriptsize 0.208}
\\
$\scriptstyle \nu_\tau\;\bar{\nu}_\tau$ & $\scriptstyle\rightarrow$ &
 $\scriptstyle W^+\;W^-$ & $\scriptstyle 639$ 
 & {\scriptsize 1.06} & {\scriptsize 1.06} & {\scriptsize 1.06} & {\scriptsize 1.06}
 & {\scriptsize 1.06} & {\scriptsize 1.06} & {\scriptsize 1.06} & {\scriptsize 1.06}
 & {\scriptsize 1.06} & {\scriptsize 1.06}
\\\hline
$\scriptstyle \nu_\mu\;\bar{\nu}_\tau$ & $\scriptstyle\rightarrow$ &
 $\scriptstyle \mu^-\;\tau^+$ & $\scriptstyle 200$ 
 & {\scriptsize 16.5} & {\scriptsize 16.5} & {\scriptsize 16.5} & {\scriptsize 16.5}
 & {\scriptsize 16.5} & {\scriptsize 16.5} & {\scriptsize 16.5} & {\scriptsize 16.5}
 & {\scriptsize 16.5} & {\scriptsize 16.5}
\\
$\scriptstyle \nu_\mu\;\bar{\nu}_\mu$ & $\scriptstyle\rightarrow$ &
 $\scriptstyle \mu^-\;\mu^+$ & $\scriptstyle 200$ 
 & {\scriptsize 22.4} & {\scriptsize 22.4} & {\scriptsize 22.4} & {\scriptsize 22.4}
 & {\scriptsize 22.4} & {\scriptsize 22.4} & {\scriptsize 22.4} & {\scriptsize 22.4}
 & {\scriptsize 22.4} & {\scriptsize 22.4}
\\
$\scriptstyle t\;\bar{t}$ & $\scriptstyle\rightarrow$ & $\scriptstyle s\;\bar{s}$ &
 $\scriptstyle 1395$ 
 & {\scriptsize 1.14} & {\scriptsize 1.14} & {\scriptsize 1.14} & {\scriptsize 1.14}
 & {\scriptsize 1.14} & {\scriptsize 1.14} & {\scriptsize 1.14} & {\scriptsize 1.14}
 & {\scriptsize 1.14} & {\scriptsize 1.14}
\\
$\scriptstyle e^-\;e^+$ & $\scriptstyle\rightarrow$ & $\scriptstyle t\;\bar{t}$ &
 $\scriptstyle 1394$ 
 & {\scriptsize 0.0723} & {\scriptsize 0.0723} & {\scriptsize 0.0724}
 & {\scriptsize 0.0723} & {\scriptsize 0.0723} & {\scriptsize 0.0723}
 & {\scriptsize 0.0723} & {\scriptsize 0.0723} & {\scriptsize 0.0723}
 & {\scriptsize 0.0723}
\\
$\scriptstyle c\;c$ & $\scriptstyle\rightarrow$ & $\scriptstyle c\;c$ &
 $\scriptstyle 200$ 
 & {\scriptsize 3.41e03} & {\scriptsize 3.41e03} & {\scriptsize 3.42e03}
 & {\scriptsize 3.41e03} & {\scriptsize 3.41e03} & {\scriptsize 3.42e03}
 & {\scriptsize 3.41e03} & {\scriptsize 3.41e03} & {\scriptsize 3.41e03}
 & {\scriptsize 3.42e03}
\\
$\scriptstyle u\;\bar{u}$ & $\scriptstyle\rightarrow$ & $\scriptstyle s\;\bar{s}$ &
 $\scriptstyle 200$ 
 & {\scriptsize 80.3} & {\scriptsize 80.3} & {\scriptsize 80.2} & {\scriptsize 80.3}
 & {\scriptsize 80.3} & {\scriptsize 80.3} & {\scriptsize 80.3} & {\scriptsize 80.3}
 & {\scriptsize 80.3} & {\scriptsize 80.3}
\\
$\scriptstyle u\;\bar{u}$ & $\scriptstyle\rightarrow$ & $\scriptstyle d\;\bar{d}$ &
 $\scriptstyle 200$ 
 & {\scriptsize 68.1} & {\scriptsize 68.1} & {\scriptsize 68.0} & {\scriptsize 68.1}
 & {\scriptsize 68.1} & {\scriptsize 68.1} & {\scriptsize 68.1} & {\scriptsize 68.1}
 & {\scriptsize 68.1} & {\scriptsize 68.0}
\\
$\scriptstyle u\;\bar{t}$ & $\scriptstyle\rightarrow$ & $\scriptstyle d\;\bar{b}$ &
 $\scriptstyle 716$ 
 & {\scriptsize 5.49} & {\scriptsize 5.49} & {\scriptsize 5.49} & {\scriptsize 5.49}
 & {\scriptsize 5.49} & {\scriptsize 5.49} & {\scriptsize 5.49} & {\scriptsize 5.49}
 & {\scriptsize 5.50} & {\scriptsize 5.49}
\\
$\scriptstyle \nu_\tau\;\tau^+$ & $\scriptstyle\rightarrow$ & $\scriptstyle u\;\bar{s}$ &
 $\scriptstyle 200$ 
 & {\scriptsize 0.501} & {\scriptsize 0.501} & {\scriptsize 0.501} & {\scriptsize 0.501}
 & {\scriptsize 0.501} & {\scriptsize 0.501} & {\scriptsize 0.501} & {\scriptsize 0.501}
 & {\scriptsize 0.501} & {\scriptsize 0.501}
\\
$\scriptstyle \nu_\mu\;\mu^+$ & $\scriptstyle\rightarrow$ & $\scriptstyle c\;\bar{s}$ &
 $\scriptstyle 200$ 
 & {\scriptsize 9.33} & {\scriptsize 9.33} & {\scriptsize 9.32} & {\scriptsize 9.33}
 & {\scriptsize 9.33} & {\scriptsize 9.33} & {\scriptsize 9.33} & {\scriptsize 9.33}
 & {\scriptsize 9.33} & {\scriptsize 9.33}
\\\hline
$\scriptstyle Z\;Z$ & $\scriptstyle\rightarrow$ & $\scriptstyle H\;H$ &
 $\scriptstyle 1690$ 
 & {\scriptsize 0.250} & {\scriptsize 0.250} & {\scriptsize 0.250} & {\scriptsize 0.250}
 & {\scriptsize 0.250} & {\scriptsize 0.250} & {\scriptsize 0.250} & {\scriptsize 0.250}
 & {\scriptsize 0.250} & {\scriptsize 0.250}
\\
$\scriptstyle W^+\;W^-$ & $\scriptstyle\rightarrow$ & $\scriptstyle H\;H$ &
 $\scriptstyle 1599$ 
 & {\scriptsize 0.151} & {\scriptsize 0.151} & {\scriptsize 0.151} & {\scriptsize 0.151}
 & {\scriptsize 0.151} & {\scriptsize 0.151} & {\scriptsize 0.151} & {\scriptsize 0.151}
 & {\scriptsize 0.151} & {\scriptsize 0.151}
\\
$\scriptstyle t\;\bar{t}$ & $\scriptstyle\rightarrow$ & $\scriptstyle H\;H$ &
 $\scriptstyle 2354$ 
 & {\scriptsize 0.0593} & {\scriptsize 0.0593} & {\scriptsize 0.0593}
 & {\scriptsize 0.0593} & {\scriptsize 0.0593} & {\scriptsize 0.0593}
 & {\scriptsize 0.0593} & {\scriptsize 0.0593} & {\scriptsize 0.0593}
 & {\scriptsize 0.0593}
\\
$\scriptstyle \tau^-\;\tau^+$ & $\scriptstyle\rightarrow$ & $\scriptstyle H\;H$ &
 $\scriptstyle 974$ 
 & {\scriptsize 2.96e-06} & {\scriptsize 2.96e-06} & {\scriptsize 2.96e-06}
 & {\scriptsize 2.96e-06} & {\scriptsize 2.96e-06} & {\scriptsize 2.96e-06}
 & {\scriptsize 2.96e-06} & {\scriptsize 2.96e-06} & {\scriptsize 2.96e-06}
 & {\scriptsize 2.96e-06}
\\
$\scriptstyle b\;\bar{b}$ & $\scriptstyle\rightarrow$ & $\scriptstyle H\;H$ &
 $\scriptstyle 998$ 
 & {\scriptsize 6.33e-06} & {\scriptsize 6.33e-06} & {\scriptsize 6.33e-06}
 & {\scriptsize 6.33e-06} & {\scriptsize 6.33e-06} & {\scriptsize 6.33e-06}
 & {\scriptsize 6.33e-06} & {\scriptsize 6.33e-06} & {\scriptsize 6.33e-06}
 & {\scriptsize 6.33e-06}
\\\hline
\end{tabular}
}
\caption{\label{tab:SM-xs}Cross sections for a selection of two-to-two processes in 
the SM. The results for {\sc CalcHep}, {\sc MadGraph}
/{\sc MadEvent}, {\sc Whizard} 1.x and {\sc Whizard} 2.0 are labeled by CH, MG,
 WO1 and WO2, respectively.
 `F' and `U' refer to the choice of the gauge in which the calculation was performed, Feyman or unitarity.
 The $p_T$ cuts were fixed according to Eq.~(\ref{eq_app:ptcut}), and the unit for
 all cross sections is $\unit{pb}$.}
\end{table*}

\begin{table*}[!h]
\centerline{\begin{tabular}{|lcl|l|lllllll|l|}
\hline & & & & \multicolumn{7}{|c|}{\sc FeynRules} & \multicolumn{1}{|c|}{Stock}\\
\multicolumn{3}{|c|}{Process} & \multicolumn{1}{|c|}{$\sqrt{s}$}
 & \multicolumn{1}{|c}{MG} & \multicolumn{1}{c}{CH} & \multicolumn{1}{c}{WO2}
 & \multicolumn{1}{c}{WO1} & \multicolumn{1}{c}{WO1} & \multicolumn{1}{c}{CH}
 & \multicolumn{1}{c|}{WO2} & \multicolumn{1}{|c|}{WO2}\\
& & & \multicolumn{1}{c}{\scriptsize[GeV]} & \multicolumn{1}{|c}{U}
 & \multicolumn{1}{c}{U} & \multicolumn{1}{c}{U} & \multicolumn{1}{c}{F}
 & \multicolumn{1}{c}{U} & \multicolumn{1}{c}{F} & \multicolumn{1}{c|}{F}
 & \multicolumn{1}{|c|}{U}\\
\hline\hline
$\scriptstyle g\;g$ & $\scriptstyle\rightarrow$ & $\scriptstyle g\;g$ &
 $\scriptstyle 200$ 
 & {\scriptsize 1.88e04} & {\scriptsize 1.88e04} & {\scriptsize 1.88e04}
 & {\scriptsize 1.88e04} & {\scriptsize 1.88e04} & {\scriptsize 1.88e04}
 & {\scriptsize 1.88e04} & {\scriptsize 1.88e04}
\\
$\scriptstyle Z\;W^+$ & $\scriptstyle\rightarrow$ &
 $\scriptstyle {Z}^\prime\;{W^+}^\prime$ & $\scriptstyle 4693$ 
 & {\scriptsize 103} & {\scriptsize 103} & {\scriptsize 103} & {\scriptsize 103}
 & {\scriptsize 103} & {\scriptsize 103} & {\scriptsize 103} & {\scriptsize 103}
\\
$\scriptstyle \gamma\;\gamma$ & $\scriptstyle\rightarrow$ & $\scriptstyle W^+\;W^-$ &
 $\scriptstyle 643$ 
 & {\scriptsize 16.1} & {\scriptsize 16.1} & {\scriptsize 16.1} & {\scriptsize 16.1}
 & {\scriptsize 16.1} & {\scriptsize 16.1} & {\scriptsize 16.1} & {\scriptsize 16.1}
\\
$\scriptstyle Z\;W^+$ & $\scriptstyle\rightarrow$ & $\scriptstyle W^+\;{Z}^\prime$ &
 $\scriptstyle 3015$ 
 & {\scriptsize 0.755} & {\scriptsize 0.755} & {\scriptsize 0.755} & {\scriptsize 0.755}
 & {\scriptsize 0.755} & {\scriptsize 0.755} & {\scriptsize 0.755} & {\scriptsize 0.756}
\\
$\scriptstyle \gamma\;Z$ & $\scriptstyle\rightarrow$ &
 $\scriptstyle {W^+}^\prime\;{W^-}^\prime$ & $\scriptstyle 8730$ 
 & {\scriptsize 0.0412} & {\scriptsize 0.0412} & {\scriptsize 0.0411}
 & {\scriptsize 0.0412} & {\scriptsize 0.0412} & {\scriptsize 0.0412}
 & {\scriptsize 0.0412} & {\scriptsize 0.0412}
\\
$\scriptstyle \gamma\;Z$ & $\scriptstyle\rightarrow$ & $\scriptstyle W^+\;{W^-}^\prime$ &
 $\scriptstyle 2686$ 
 & {\scriptsize 1.30} & {\scriptsize 1.30} & {\scriptsize 1.30} & {\scriptsize 1.30}
 & {\scriptsize 1.30} & {\scriptsize 1.30} & {\scriptsize 1.30} & {\scriptsize 1.30}
\\
$\scriptstyle W^+\;{W^+}^\prime$ & $\scriptstyle\rightarrow$ &
 $\scriptstyle {W^+}^\prime\;{W^+}^\prime$ & $\scriptstyle 6322$ 
 & {\scriptsize 3.32} & {\scriptsize 3.31} & {\scriptsize 3.31} & {\scriptsize 3.31}
 & {\scriptsize 3.31} & {\scriptsize 3.31} & {\scriptsize 3.31} & {\scriptsize 3.32}
\\
$\scriptstyle Z\;{Z}^\prime$ & $\scriptstyle\rightarrow$ &
 $\scriptstyle {W^+}^\prime\;{W^-}^\prime$ & $\scriptstyle 6371$ 
 & {\scriptsize 3.86} & {\scriptsize 3.86} & {\scriptsize 3.86} & {\scriptsize 3.86}
 & {\scriptsize 3.86} & {\scriptsize 3.86} & {\scriptsize 3.86} & {\scriptsize 3.86}
\\
$\scriptstyle W^+\;W^+$ & $\scriptstyle\rightarrow$ & $\scriptstyle W^+\;{W^+}^\prime$ &
 $\scriptstyle 2965$ 
 & {\scriptsize 2.41} & {\scriptsize 2.42} & {\scriptsize 2.42} & {\scriptsize 2.42}
 & {\scriptsize 2.42} & {\scriptsize 2.42} & {\scriptsize 2.42} & {\scriptsize 2.42}
\\
$\scriptstyle \gamma\;W^+$ & $\scriptstyle\rightarrow$ &
 $\scriptstyle {Z}^\prime\;{W^+}^\prime$ & $\scriptstyle 8656$ 
 & {\scriptsize 0.0349} & {\scriptsize 0.0349} & {\scriptsize 0.0349}
 & {\scriptsize 0.0349} & {\scriptsize 0.0349} & {\scriptsize 0.0349}
 & {\scriptsize 0.0349} & {\scriptsize 0.0349}
\\\hline
$\scriptstyle W^+\;{W^-}^\prime$ & $\scriptstyle\rightarrow$ &
 $\scriptstyle s\;{\bar{s}}^\prime$ & $\scriptstyle 18765$ 
 & {\scriptsize 33.6} & {\scriptsize 33.6} & {\scriptsize 33.6} & {\scriptsize 33.6}
 & {\scriptsize 33.6} & {\scriptsize 33.6} & {\scriptsize 33.6} & {\scriptsize 33.6}
\\
$\scriptstyle {Z}^\prime\;{Z}^\prime$ & $\scriptstyle\rightarrow$ &
 $\scriptstyle b\;\bar{b}$ & $\scriptstyle 8094$ 
 & {\scriptsize 0.0958} & {\scriptsize 0.0959} & {\scriptsize 0.0959}
 & {\scriptsize 0.0959} & {\scriptsize 0.0959} & {\scriptsize 0.0959}
 & {\scriptsize 0.0959} & {\scriptsize 0.0959}
\\
$\scriptstyle {Z}^\prime\;{Z}^\prime$ & $\scriptstyle\rightarrow$ &
 $\scriptstyle {e^-}^\prime\;{e^+}^\prime$ & $\scriptstyle 36900$ 
 & {\scriptsize 1.54} & {\scriptsize 1.54} & {\scriptsize 1.54} & {\scriptsize 1.54}
 & {\scriptsize 1.54} & {\scriptsize 1.54} & {\scriptsize 1.54} & {\scriptsize 1.54}
\\
$\scriptstyle \gamma\;{W^+}^\prime$ & $\scriptstyle\rightarrow$ &
 $\scriptstyle {c}^\prime\;{\bar{s}}^\prime$ & $\scriptstyle 34886$ 
 & {\scriptsize 0.00220} & {\scriptsize 0.00220} & {\scriptsize 0.00220}
 & {\scriptsize 0.00220} & {\scriptsize 0.00220} & {\scriptsize 0.00220}
 & {\scriptsize 0.00220} & {\scriptsize 0.00220}
\\
$\scriptstyle W^+\;{W^-}^\prime$ & $\scriptstyle\rightarrow$ &
 $\scriptstyle {\tau^-}^\prime\;{\tau^+}^\prime$ & $\scriptstyle 35208$ 
 & {\scriptsize 1.45} & {\scriptsize 1.45} & {\scriptsize 1.45} & {\scriptsize 1.45}
 & {\scriptsize 1.45} & {\scriptsize 1.45} & {\scriptsize 1.45} & {\scriptsize 1.45}
\\
$\scriptstyle W^+\;{W^-}^\prime$ & $\scriptstyle\rightarrow$ &
 $\scriptstyle u\;{\bar{u}}^\prime$ & $\scriptstyle 18765$ 
 & {\scriptsize 34.0} & {\scriptsize 34.0} & {\scriptsize 34.0} & {\scriptsize 34.0}
 & {\scriptsize 34.0} & {\scriptsize 34.0} & {\scriptsize 34.0} & {\scriptsize 34.0}
\\
$\scriptstyle Z\;Z$ & $\scriptstyle\rightarrow$ & $\scriptstyle u\;{\bar{u}}^\prime$ &
 $\scriptstyle 17173$ 
 & {\scriptsize 30.3} & {\scriptsize 30.3} & {\scriptsize 30.3} & {\scriptsize 30.3}
 & {\scriptsize 30.3} & {\scriptsize 30.3} & {\scriptsize 30.3} & {\scriptsize 30.3}
\\
$\scriptstyle {W^+}^\prime\;{W^-}^\prime$ & $\scriptstyle\rightarrow$ &
 $\scriptstyle b\;{\bar{b}}^\prime$ & $\scriptstyle 20460$ 
 & {\scriptsize 38.5} & {\scriptsize 38.5} & {\scriptsize 38.5} & {\scriptsize 38.5}
 & {\scriptsize 38.5} & {\scriptsize 38.5} & {\scriptsize 38.5} & {\scriptsize 38.4}
\\
$\scriptstyle {W^+}^\prime\;{W^-}^\prime$ & $\scriptstyle\rightarrow$ &
 $\scriptstyle {\nu_e}^\prime\;{\bar{\nu}_e}^\prime$ & $\scriptstyle 36886$ 
 & {\scriptsize 1.64} & {\scriptsize 1.64} & {\scriptsize 1.64} & {\scriptsize 1.64}
 & {\scriptsize 1.64} & {\scriptsize 1.64} & {\scriptsize 1.64} & {\scriptsize 1.64}
\\
$\scriptstyle \gamma\;\gamma$ & $\scriptstyle\rightarrow$ &
 $\scriptstyle {\mu^-}^\prime\;{\mu^+}^\prime$ & $\scriptstyle 32886$ 
 & {\scriptsize 0.000463} & {\scriptsize 0.000463} & {\scriptsize 0.000463}
 & {\scriptsize 0.000463} & {\scriptsize 0.000463} & {\scriptsize 0.000463}
 & {\scriptsize 0.000463} & {\scriptsize 0.000463}
\\\hline
$\scriptstyle \nu_\tau\;u$ & $\scriptstyle\rightarrow$ &
 $\scriptstyle {\nu_\tau}^\prime\;{u}^\prime$ & $\scriptstyle 32886$ 
 & {\scriptsize 0.0966} & {\scriptsize 0.0966} & {\scriptsize 0.0967}
 & {\scriptsize 0.0966} & {\scriptsize 0.0966} & {\scriptsize 0.0966}
 & {\scriptsize 0.0966} & {\scriptsize 0.0966}
\\
$\scriptstyle \nu_e\;u$ & $\scriptstyle\rightarrow$ &
 $\scriptstyle {\nu_e}^\prime\;{u}^\prime$ & $\scriptstyle 32886$ 
 & {\scriptsize 0.0966} & {\scriptsize 0.0966} & {\scriptsize 0.0966}
 & {\scriptsize 0.0966} & {\scriptsize 0.0966} & {\scriptsize 0.0966}
 & {\scriptsize 0.0967} & {\scriptsize 0.0967}
\\
$\scriptstyle c\;\bar{c}$ & $\scriptstyle\rightarrow$ &
 $\scriptstyle {\nu_e}^\prime\;{\bar{\nu}_e}^\prime$ & $\scriptstyle 32896$ 
 & {\scriptsize 5.38e-06} & {\scriptsize 5.38e-06} & {\scriptsize 5.38e-06}
 & {\scriptsize 5.38e-06} & {\scriptsize 5.38e-06} & {\scriptsize 5.38e-06}
 & {\scriptsize 5.38e-06} & {\scriptsize 5.38e-06}
\\
$\scriptstyle b\;{u}^\prime$ & $\scriptstyle\rightarrow$ &
 $\scriptstyle {t}^\prime\;{d}^\prime$ & $\scriptstyle 49606$ 
 & {\scriptsize 0.0337} & {\scriptsize 0.0337} & {\scriptsize 0.0337}
 & {\scriptsize 0.0337} & {\scriptsize 0.0337} & {\scriptsize 0.0337}
 & {\scriptsize 0.0337} & {\scriptsize 0.0337}
\\
$\scriptstyle \tau^-\;\bar{d}$ & $\scriptstyle\rightarrow$ &
 $\scriptstyle {\nu_\tau}^\prime\;{\bar{u}}^\prime$ & $\scriptstyle 32893$ 
 & {\scriptsize 0.386} & {\scriptsize 0.386} & {\scriptsize 0.386} & {\scriptsize 0.386}
 & {\scriptsize 0.386} & {\scriptsize 0.386} & {\scriptsize 0.386} & {\scriptsize 0.386}
\\
$\scriptstyle e^-\;b$ & $\scriptstyle\rightarrow$ &
 $\scriptstyle {e^-}^\prime\;{b}^\prime$ & $\scriptstyle 32903$ 
 & {\scriptsize 0.0965} & {\scriptsize 0.0965} & {\scriptsize 0.0965}
 & {\scriptsize 0.0965} & {\scriptsize 0.0965} & {\scriptsize 0.0965}
 & {\scriptsize 0.0966} & {\scriptsize 0.0965}
\\
$\scriptstyle t\;{\bar{\nu}_\tau}^\prime$ & $\scriptstyle\rightarrow$ &
 $\scriptstyle {\tau^+}^\prime\;{b}^\prime$ & $\scriptstyle 50014$ 
 & {\scriptsize 0.0430} & {\scriptsize 0.0429} & {\scriptsize 0.0429}
 & {\scriptsize 0.0429} & {\scriptsize 0.0429} & {\scriptsize 0.0429}
 & {\scriptsize 0.0429} & {\scriptsize 0.0429}
\\
$\scriptstyle c\;\bar{s}$ & $\scriptstyle\rightarrow$ &
 $\scriptstyle {\nu_e}^\prime\;{e^+}^\prime$ & $\scriptstyle 32892$ 
 & {\scriptsize 1.82e-05} & {\scriptsize 1.82e-05} & {\scriptsize 1.82e-05}
 & {\scriptsize 1.82e-05} & {\scriptsize 1.82e-05} & {\scriptsize 1.82e-05}
 & {\scriptsize 1.82e-05} & {\scriptsize 1.82e-05}
\\
$\scriptstyle \nu_\tau\;\bar{\nu}_\tau$ & $\scriptstyle\rightarrow$ &
 $\scriptstyle t\;{\bar{t}}^\prime$ & $\scriptstyle 17388$ 
 & {\scriptsize 7.23e-06} & {\scriptsize 7.24e-06} & {\scriptsize 7.24e-06}
 & {\scriptsize 7.24e-06} & {\scriptsize 7.24e-06} & {\scriptsize 7.24e-06}
 & {\scriptsize 7.24e-06} & {\scriptsize 7.24e-06}
\\
$\scriptstyle \tau^-\;\tau^+$ & $\scriptstyle\rightarrow$ &
 $\scriptstyle s\;{\bar{s}}^\prime$ & $\scriptstyle 16458$ 
 & {\scriptsize 9.33e-06} & {\scriptsize 9.35e-06} & {\scriptsize 9.35e-06}
 & {\scriptsize 9.35e-06} & {\scriptsize 9.35e-06} & {\scriptsize 9.35e-06}
 & {\scriptsize 9.35e-06} & {\scriptsize 9.35e-06}
\\\hline
\end{tabular}
}
\caption{\label{tab:MHM-xs}Cross sections for a selection of two-to-two processes in 
the MHM. The results for {\sc CalcHep}, {\sc MadGraph}
/{\sc MadEvent}, {\sc Whizard} 1.x and {\sc Whizard} 2.0 are denoted by CH, MG,
 WO1 and WO2, respectively. `F' and `U' refer to the choice of the gauge in which the
 calculation was performed, Feyman or unitarity.
  The $p_T$ cuts were fixed according to Eq.~(\ref{eq_app:ptcut}), and
  the unit for all cross sections is $\unit{pb}$.
  A trailing prime
  ``$\prime$'' at the end of a particle identifier denotes its heavy partner
  (\emph{cf.} Ref. \cite{Chivukula:2006cg}).}
\end{table*}

\begin{table*}[!h]
\centerline{\begin{tabular}{|lcl|l|llll|ll|}
\hline & & & & \multicolumn{4}{|c|}{\sc FeynRules} & \multicolumn{2}{|c|}{Stock}\\
\multicolumn{3}{|c|}{Process} & \multicolumn{1}{|c|}{$\sqrt{s}$}
 & \multicolumn{1}{|c}{CH} & \multicolumn{1}{c}{MG} & \multicolumn{1}{c}{WO1}
 & \multicolumn{1}{c|}{WO2} & \multicolumn{1}{|c}{MG} & \multicolumn{1}{c|}{WO2}\\
& & & \multicolumn{1}{c}{\scriptsize[GeV]} & \multicolumn{1}{|c}{U}
 & \multicolumn{1}{c}{U} & \multicolumn{1}{c}{U} & \multicolumn{1}{c|}{U}
 & \multicolumn{1}{|c}{U} & \multicolumn{1}{c|}{U}\\
\hline\hline
$\scriptstyle Z\;Z$ & $\scriptstyle\rightarrow$ & $\scriptstyle W^+\;W^-$ &
 $\scriptstyle 1368$ 
 & {\scriptsize 26.2} & {\scriptsize 26.3} & {\scriptsize 26.2} & {\scriptsize 26.2}
 & {\scriptsize 26.3} & {\scriptsize 26.2}
\\
$\scriptstyle W^+\;W^+$ & $\scriptstyle\rightarrow$ & $\scriptstyle W^+\;W^+$ &
 $\scriptstyle 1277$ 
 & {\scriptsize 25.6} & {\scriptsize 25.7} & {\scriptsize 25.6} & {\scriptsize 25.7}
 & {\scriptsize 25.7} & {\scriptsize 25.6}
\\
$\scriptstyle \gamma\;Z$ & $\scriptstyle\rightarrow$ & $\scriptstyle W^+\;W^-$ &
 $\scriptstyle 1003$ 
 & {\scriptsize 19.3} & {\scriptsize 19.3} & {\scriptsize 19.3} & {\scriptsize 19.3}
 & {\scriptsize 19.3} & {\scriptsize 19.3}
\\
$\scriptstyle g\;g$ & $\scriptstyle\rightarrow$ & $\scriptstyle g\;g$ &
 $\scriptstyle 800$ 
 & {\scriptsize 839} & {\scriptsize 840} & {\scriptsize 839} & {\scriptsize 839}
 & {\scriptsize 839} & {\scriptsize 839}
\\
$\scriptstyle Z\;Z$ & $\scriptstyle\rightarrow$ & $\scriptstyle Z\;Z$ &
 $\scriptstyle 1459$ 
 & {\scriptsize 0.223} & {\scriptsize 0.223} & {\scriptsize 0.223} & {\scriptsize 0.223}
 & {\scriptsize 0.223} & {\scriptsize 0.223}
\\
$\scriptstyle \gamma\;\gamma$ & $\scriptstyle\rightarrow$ & $\scriptstyle W^+\;W^-$ &
 $\scriptstyle 1278$ 
 & {\scriptsize 4.95} & {\scriptsize 4.96} & {\scriptsize 4.95} & {\scriptsize 4.95}
 & {\scriptsize 4.96} & {\scriptsize 4.95}
\\\hline
$\scriptstyle \widetilde{g}\;\widetilde{g}$ & $\scriptstyle\rightarrow$ &
 $\scriptstyle g\;g$ & $\scriptstyle 4862$ 
 & {\scriptsize 0.977} & {\scriptsize 0.977} & {\scriptsize 0.977} & {\scriptsize 0.977}
 & {\scriptsize 0.977} & {\scriptsize 0.977}
\\
$\scriptstyle \tau^-\;\tau^+$ & $\scriptstyle\rightarrow$ & $\scriptstyle \gamma\;Z$ &
 $\scriptstyle 758$ 
 & {\scriptsize 0.327} & {\scriptsize 0.327} & {\scriptsize 0.327} & {\scriptsize 0.327}
 & {\scriptsize 0.327} & {\scriptsize 0.327}
\\
$\scriptstyle \widetilde{\chi}^0_4\;\widetilde{\chi}_1^+$ & $\scriptstyle\rightarrow$ &
 $\scriptstyle \gamma\;W^+$ & $\scriptstyle 2573$ 
 & {\scriptsize 0.0219} & {\scriptsize 0.0219} & {\scriptsize 0.0219}
 & {\scriptsize 0.0219} & {\scriptsize 0.0219} & {\scriptsize 0.0219}
\\
$\scriptstyle d\;\bar{d}$ & $\scriptstyle\rightarrow$ & $\scriptstyle Z\;Z$ &
 $\scriptstyle 730$ 
 & {\scriptsize 0.0838} & {\scriptsize 0.0839} & {\scriptsize 0.0838}
 & {\scriptsize 0.0838} & {\scriptsize 0.0839} & {\scriptsize 0.0839}
\\
$\scriptstyle t\;\bar{t}$ & $\scriptstyle\rightarrow$ & $\scriptstyle g\;g$ &
 $\scriptstyle 1400$ 
 & {\scriptsize 6.00} & {\scriptsize 6.00} & {\scriptsize 6.00} & {\scriptsize 5.99}
 & {\scriptsize 6.00} & {\scriptsize 5.99}
\\
$\scriptstyle c\;\bar{c}$ & $\scriptstyle\rightarrow$ & $\scriptstyle \gamma\;\gamma$ &
 $\scriptstyle 800$ 
 & {\scriptsize 0.0272} & {\scriptsize 0.0272} & {\scriptsize 0.0272}
 & {\scriptsize 0.0272} & {\scriptsize 0.0272} & {\scriptsize 0.0272}
\\
$\scriptstyle b\;\bar{b}$ & $\scriptstyle\rightarrow$ & $\scriptstyle W^+\;W^-$ &
 $\scriptstyle 678$ 
 & {\scriptsize 1.72} & {\scriptsize 1.72} & {\scriptsize 1.72} & {\scriptsize 1.72}
 & {\scriptsize 1.72} & {\scriptsize 1.72}
\\
$\scriptstyle \widetilde{\chi}^0_1\;\widetilde{\chi}^0_1$ & $\scriptstyle\rightarrow$ &
 $\scriptstyle W^+\;W^-$ & $\scriptstyle 1412$ 
 & {\scriptsize 0.00586} & {\scriptsize 0.00585} & {\scriptsize 0.00586}
 & {\scriptsize 0.00586} & {\scriptsize 0.00587} & {\scriptsize 0.00586}
\\\hline
$\scriptstyle c\;\bar{s}$ & $\scriptstyle\rightarrow$ &
 $\scriptstyle \widetilde{\chi}^0_1\;\widetilde{\chi}_1^+$ & $\scriptstyle 1114$ 
 & {\scriptsize 0.000873} & {\scriptsize 0.000874} & {\scriptsize 0.000873}
 & {\scriptsize 0.000873} & {\scriptsize 0.000874} & {\scriptsize 0.000873}
\\
$\scriptstyle \widetilde{\chi}^0_2\;\widetilde{\chi}^0_2$ & $\scriptstyle\rightarrow$ &
 $\scriptstyle \widetilde{\chi}^0_2\;\widetilde{\chi}^0_2$ & $\scriptstyle 2897$ 
 & {\scriptsize 0.00237} & {\scriptsize 0.00237} & {\scriptsize 0.00237}
 & {\scriptsize 0.00237} & {\scriptsize 0.00236} & {\scriptsize 0.00237}
\\
$\scriptstyle \widetilde{\chi}_1^+\;\widetilde{\chi}_2^+$ & $\scriptstyle\rightarrow$ &
 $\scriptstyle \widetilde{\chi}_2^+\;\widetilde{\chi}_2^+$ & $\scriptstyle 5286$ 
 & {\scriptsize 0.0151} & {\scriptsize 0.0151} & {\scriptsize 0.0151}
 & {\scriptsize 0.0151} & {\scriptsize 0.0151} & {\scriptsize 0.0151}
\\
$\scriptstyle c\;\bar{c}$ & $\scriptstyle\rightarrow$ &
 $\scriptstyle \widetilde{\chi}^0_2\;\widetilde{\chi}^0_3$ & $\scriptstyle 2179$ 
 & {\scriptsize 0.000134} & {\scriptsize 0.000134} & {\scriptsize 0.000134}
 & {\scriptsize 0.000134} & {\scriptsize 0.000134} & {\scriptsize 0.000134}
\\
$\scriptstyle s\;\bar{s}$ & $\scriptstyle\rightarrow$ &
 $\scriptstyle \widetilde{g}\;\widetilde{g}$ & $\scriptstyle 4862$ 
 & {\scriptsize 0.144} & {\scriptsize 0.144} & {\scriptsize 0.144} & {\scriptsize 0.144}
 & {\scriptsize 0.144} & {\scriptsize 0.144}
\\
$\scriptstyle \mu^-\;\mu^+$ & $\scriptstyle\rightarrow$ &
 $\scriptstyle \widetilde{\chi}^0_2\;\widetilde{\chi}^0_2$ & $\scriptstyle 1449$ 
 & {\scriptsize 0.0291} & {\scriptsize 0.0291} & {\scriptsize 0.0291}
 & {\scriptsize 0.0291} & {\scriptsize 0.0291} & {\scriptsize 0.0291}
\\
$\scriptstyle t\;\bar{b}$ & $\scriptstyle\rightarrow$ &
 $\scriptstyle \widetilde{\chi}^0_3\;\widetilde{\chi}_2^+$ & $\scriptstyle 3694$ 
 & {\scriptsize 0.00759} & {\scriptsize 0.00759} & {\scriptsize 0.00758}
 & {\scriptsize 0.00759} & {\scriptsize 0.00759} & {\scriptsize 0.00759}
\\
$\scriptstyle c\;\bar{c}$ & $\scriptstyle\rightarrow$ &
 $\scriptstyle \widetilde{g}\;\widetilde{g}$ & $\scriptstyle 4862$ 
 & {\scriptsize 0.144} & {\scriptsize 0.144} & {\scriptsize 0.144} & {\scriptsize 0.144}
 & {\scriptsize 0.144} & {\scriptsize 0.144}
\\\hline
$\scriptstyle \tau^-\;u$ & $\scriptstyle\rightarrow$ &
 $\scriptstyle \tilde{l}_6^-\;\tilde{u}_5$ & $\scriptstyle 3079$ 
 & {\scriptsize 0.000510} & {\scriptsize 0.000510} & {\scriptsize 0.000510}
 & {\scriptsize 0.000510} & {\scriptsize 0.000510} & {\scriptsize 0.000510}
\\
$\scriptstyle \nu_\mu\;\nu_\tau$ & $\scriptstyle\rightarrow$ &
 $\scriptstyle \tilde{\nu}_1\;\tilde{\nu}_2$ & $\scriptstyle 1480$ 
 & {\scriptsize 0.00902} & {\scriptsize 0.00903} & {\scriptsize 0.00902}
 & {\scriptsize 0.00902} & {\scriptsize 0.00903} & {\scriptsize 0.00902}
\\
$\scriptstyle \gamma\;Z$ & $\scriptstyle\rightarrow$ &
 $\scriptstyle \tilde{l}_1^-\;\tilde{l}_6^+$ & $\scriptstyle 1730$ 
 & {\scriptsize 0.00278} & {\scriptsize 0.00278} & {\scriptsize 0.00278}
 & {\scriptsize 0.00278} & {\scriptsize 0.00278} & {\scriptsize 0.00278}
\\
$\scriptstyle s\;\widetilde{\chi}^0_2$ & $\scriptstyle\rightarrow$ &
 $\scriptstyle \tilde{d}_3\;H_1$ & $\scriptstyle 3349$ 
 & {\scriptsize 9.85e-06} & {\scriptsize 9.85e-06} & {\scriptsize 9.85e-06}
 & {\scriptsize 9.85e-06} & {\scriptsize 9.85e-06} & {\scriptsize 9.85e-06}
\\
$\scriptstyle \gamma\;Z$ & $\scriptstyle\rightarrow$ &
 $\scriptstyle \tilde{u}_5\;\tilde{u}^*_5$ & $\scriptstyle 4854$ 
 & {\scriptsize 0.00268} & {\scriptsize 0.00268} & {\scriptsize 0.00268}
 & {\scriptsize 0.00268} & {\scriptsize 0.00268} & {\scriptsize 0.00268}
\\
$\scriptstyle \nu_e\;e^-$ & $\scriptstyle\rightarrow$ &
 $\scriptstyle \tilde{\nu}_3\;\tilde{l}_2^-$ & $\scriptstyle 1317$ 
 & {\scriptsize 0.0274} & {\scriptsize 0.0274} & {\scriptsize 0.0274}
 & {\scriptsize 0.0274} & {\scriptsize 0.0274} & {\scriptsize 0.0274}
\\
$\scriptstyle s\;\widetilde{\chi}^0_1$ & $\scriptstyle\rightarrow$ &
 $\scriptstyle \tilde{u}_4\;H^-$ & $\scriptstyle 4263$ 
 & {\scriptsize 9.84e-06} & {\scriptsize 9.84e-06} & {\scriptsize 9.84e-06}
 & {\scriptsize 9.84e-06} & {\scriptsize 9.84e-06} & {\scriptsize 9.83e-06}
\\
$\scriptstyle \tau^-\;\tau^+$ & $\scriptstyle\rightarrow$ &
 $\scriptstyle \tilde{u}_1\;\tilde{u}^*_6$ & $\scriptstyle 3956$ 
 & {\scriptsize 0.000449} & {\scriptsize 0.000449} & {\scriptsize 0.000449}
 & {\scriptsize 0.000449} & {\scriptsize 0.000449} & {\scriptsize 0.000449}
\\\hline
\end{tabular}
}
\caption{\label{tab:MSSM-xs}Cross sections for a selection of two-to-two processes in 
the MSSM. The results for {\sc CalcHep}, {\sc MadGraph}
/{\sc MadEvent}, {\sc Whizard} 1.x and {\sc Whizard} 2.0 are denoted by CH, MG,
 WO1 and WO2, respectively.
 `F' and `U' refer to the choice of the gauge in which the
 calculation was performed, Feyman or unitarity.
  The $p_T$ cuts were fixed according to Eq.~(\ref{eq_app:ptcut}), and the unit
  for all cross sections is $\unit{pb}$. The sfermions are
  labeled according to the SLHA2 conventions (\emph{cf.} Ref.
  \cite{Allanach:2008qq}).}
\end{table*}

\clearpage

\end{appendix}

\end{document}